\documentclass[apj,twocolumn,appendixfloats]{myemulateapj}
\pdfoutput=1
\newenvironment{enumerate*}{
\vspace{-1.0mm}
\begin{enumerate}
  \setlength{\parskip}{1pt}
}{\end{enumerate}
}

\newcommand{\fref}[1]{Figure~\ref{figure:#1}}
\newcommand{\tref}[1]{Table~\ref{table:#1}}
\newcommand{\iref}[1]{Step~\ref{item:#1}}
\newcommand{\irefs}[2]{Steps~\ref{item:#1} and~\ref{item:#2}}
\newcommand{\sref}[1]{\S\ref{section:#1}}

\newcommand{\thefiguretitle}{None}
\newcommand{\figuretitle}[1]{\renewcommand{\thefiguretitle}{{\bf #1}}}
\renewcommand{\thefootnote}{\fnsymbol{footnote}}

\usepackage[pdftitle={The BOSS Emission-Line Lens Survey (BELLS). I.  A large spectroscopically selected sample of Lens Galaxies at redshift ~ 0.5},pdfauthor={Joel R. Brownstein}, pdfsubject={Strong Galaxy-Galaxy Lensing}, pdfkeywords={galaxies: distances and redshifts, galaxies: evolution, galaxies: high-redshift, gravitational lensing: strong, cosmology: large-scale structure of universe, surveys}, bookmarks, bookmarksopen, colorlinks=true, linkcolor=blue, citecolor=blue, anchorcolor=blue, urlcolor=blue]{hyperref}
\usepackage{graphicx}
\usepackage{natbib}
\usepackage{multirow}
\journalinfo{The Astrophysical Journal, \href{http://iopscience.iop.org/0004-637X/744/1/41/}{744:41, 2012 January 1}}
\doiinfo{doi:~\href{http://dx.doi.org/10.1088/0004-637X/744/1/41}{10.1088/0004-637X/744/1/41}}
\submitted{Received 2011 October 6; accepted 2011 November 16; published 2011 December 13}
\begin{document}

\title{The BOSS Emission-Line Lens Survey (BELLS).\\I. A large spectroscopically selected sample of Lens Galaxies at redshift \(\sim 0.5\)\altaffilmark{\(*\)}}
\shorttitle{BELLS I}
\shortauthors{Brownstein {\em et al.}}
\author{Joel R.\,Brownstein\altaffilmark{1}, Adam S.\, Bolton\altaffilmark{1}, David J.\,Schlegel\altaffilmark{2}, Daniel J.\,Eisenstein\altaffilmark{3}, Christopher S.\,Kochanek\altaffilmark{4}, Natalia Connolly\altaffilmark{5}, Claudia Maraston\altaffilmark{6}, Parul Pandey\altaffilmark{1,7}, Stella Seitz\altaffilmark{8}, David A.\,Wake\altaffilmark{9}, W.\,Michael~Wood-Vasey\altaffilmark{10}, Jon Brinkmann\altaffilmark{11}, Donald P. Schneider\altaffilmark{12} and Benjamin A. Weaver\altaffilmark{13}}
\affiliation{\(^1\)Department of Physics and Astronomy, University of Utah, Salt Lake City, UT 84112, USA.}
\affiliation{\(^2\)Lawrence Berkeley National Laboratory, Berkeley, CA 94720, USA.}
\affiliation{\(^3\)Harvard College Observatory, 60 Garden Street MS 20, Cambridge, MA 02138, USA.}
\affiliation{\(^4\)Department of Astronomy and Center for Cosmology and Astroparticle Physics, Ohio State University, Columbus, OH 43210, USA.}
\affiliation{\(^5\)Department of Physics, Hamilton College, Clinton, NY 13323, USA.}
\affiliation{\(^6\)Institute of Cosmology and Gravitation, University of Portsmouth, Portsmouth PO1 3FX, UK.}
\affiliation{\(^7\)Department of Electrical and Computer Engineering, Rutgers, The State University of New Jersey, Piscataway, NJ 08854, USA.}
\affiliation{\(^8\)University Observatory M\"unich, Scheinstrasse 1, 81679 Munchen, Germany} 
\affiliation{\(^9\)Department of Astronomy, Yale University, New Haven, CT 06520, USA.}
\affiliation{\(^{10}\)Pittsburgh Center for Particle Physics, Astrophysics, and Cosmology (PITT-PACC).  Department of Physics and Astronomy, University of Pittsburgh, Pittsburgh, PA 15260, USA.}
\affiliation{\(^{11}\)Apache Point Observatory, P.O.\,Box 59, Sunspot, NM 88349, USA.}
\affiliation{\(^{12}\)Department of Astronomy and Astrophysics and Institute for Gravitation and the Cosmos, The Pennsylvania State University, University Park, PA 16802, USA.}
\affiliation{\(^{13}\)Center for Cosmology and Particle Physics, New York University, New York, NY 10003, USA.}
\email{joelbrownstein@astro.utah.edu}
\email{bolton@astro.utah.edu}
\altaffiltext{\(*\)}{Based on observations made with the NASA/ESA {\em Hubble Space Telescope}, obtained at the Space Telescope Science Institute, which is operated by the Association of Universities for Research in Astronomy, Inc., under NASA contract NAS 5-26555. These observations are associated with program 12209. Based on spectroscopic data from the Baryon Oscillation Spectroscopic Survey of the Sloan Digital Sky Survey III.}
\begin{abstract}
We present a catalog of 25 definite and 11 probable strong galaxy--galaxy gravitational lens systems with lens redshifts \(0.4 \lesssim z \lesssim 0.7\), discovered spectroscopically by the presence of higher redshift emission-lines within the Baryon Oscillation Spectroscopic Survey (BOSS) of luminous galaxies, and confirmed with high-resolution \textsl{Hubble Space Telescope} (\textsl{HST}) images of 44 candidates.  Our survey  extends the methodology of the Sloan Lens ACS Survey (SLACS) to higher redshift. We describe the details of the BOSS spectroscopic candidate detections, our \textsl{HST} Advanced Camera for Surveys (ACS) image processing and analysis methods, and our strong gravitational lens modeling procedure.  We report BOSS spectroscopic parameters and ACS photometric parameters for all candidates, and mass-distribution parameters for the best-fit singular isothermal ellipsoid models of definite lenses. Our sample to date was selected using only the first six months of BOSS survey-quality spectroscopic data.  The full five-year BOSS database should produce a sample of several hundred strong galaxy--galaxy lenses and in combination with SLACS lenses at lower redshift, strongly constrain the redshift evolution of the structure of elliptical, bulge-dominated galaxies as a function of luminosity, stellar mass, and rest-frame color, thereby providing a powerful test for competing theories of galaxy formation and evolution.
\end{abstract} 
 

\section{Introduction} \label{section:introduction} Strong galaxy--galaxy lensing produces unmistakably distorted, amplified, and multiplied images of a distant galaxy through the gravitational field of a foreground-galaxy.  This phenomenon provides a direct astrometric measurement of the total mass distribution within the interior of the lens galaxy, including luminous and dark components.  Strong lensing on galactic scales was first suggested by \citet{Zwicky:1937PhRv...51..290Z} and first discovered by \citet{Walsh:1979Natur.279..381W}.  Since then, a number of dedicated and serendipitous surveys have produced a combined sample of over 200 galaxy-scale lens systems~\protect\citep[for a recent review, see][]{Treu:2010ARA&A..48...87T}.  Of the various survey methods, the most prolific means of identifying strong galaxy--galaxy lens candidates is spectroscopic discovery, which provides evidence for two galaxies along one line of sight, along with precise measurements of the  lens and source redshifts~\protect\citep{Warren:1996MNRAS.278..139W,Bolton:2004AJ....127.1860B}.  Images taken with the \textsl{Hubble Space Telescope} (\textsl{HST}) or other high-resolution facilities then provide a confirmation of the lensing hypothesis and make precise (milliarcsecond) measurements of the relative positions of the components.  This method has been employed by the Sloan Lens ACS (SLACS) survey~\protect\citep{Bolton:2006ApJ...638..703B,Bolton:2008ApJ...682..964B} to produce a sample of 85 grade-A confirmed strong lenses and an additional 13 grade-B probable lenses~\protect\citep{Auger:2009ApJ...705.1099A}.

Strong lensing is the best probe of the distribution of mass in the transition region between the baryon-dominated central regions of  galaxies and the dark matter halo~\protect\citep[e.g.,][]{Kochanek:2005IAUS..225..205K}. By combining strong-lensing aperture masses with lens-galaxy luminosities, sizes, and velocity dispersions, the SLACS collaboration has arrived at the following results~(\protect\citealt{Treu:2006ApJ...640..662T,Koopmans:2006ApJ...649..599K,Gavazzi:2007ApJ...667..176G,Bolton:2008ApJ...682..964B,Koopmans:2009ApJ...703L..51K}, see also \protect\citealt{Jiang.2007ApJ...671.1568J}).
\begin{enumerate*}
\item The average total mass-density profile of early-type galaxies is nearly isothermal, corresponding to nearly flat galaxy rotation curves within the Einstein radius.
\item Models for early-type galaxies in which light linearly traces mass are falsified at \(\gtrsim 99.9\%\) confidence even within the optical half-light radius.
\item The average weak--lensing signal out to a few hundred kiloparsecs around the SLACS lenses is consistent with an isothermal extrapolation of the strong-lensing constraints from the inner few kiloparsecs.
\item The average density structure of elliptical galaxies is universal and does not vary systematically with mass and exhibits a small but significant intrinsic scatter that is not significantly correlated with any other observable quantity.
\item The relationship between the lensing mass and the dynamical mass of SLACS lenses has a logarithmic slope of \(1.03 \pm 0.04\).
\item The variation in the dark matter fraction suggests a {\it bulge--halo conspiracy}, with the combined rotation curve of both components together being flatter than that of either component independently.
\end{enumerate*}

Many of these results were also found in earlier but smaller studies of lensed quasars~(see e.g., \citealt{Treu.2004ApJ...611..739T}, or \citealt{Rusin.2005ApJ...623..666R}, and references therein). 

In order to conduct the next generation of large-scale structure surveys,
the Sloan Digital Sky Survey III (SDSS-III) collaboration upgraded the Sloan Digital Sky Survey I (SDSS-I) optical spectrographs~\protect\citep{York.2000AJ....120.1579Y} and initiated
the Baryon Oscillation Spectroscopic Survey~\protect\citep[BOSS; ][]{Eisenstein:2011AJ....142...72E}.
The significant hardware changes relative to SDSS-I are (1) an increase in the number
of fibers from 640 to 1000, (2) smaller fiber diameters that subtend \(2\arcsec\)
rather than \(3\arcsec\) on the sky, (3) volume-phase holographic
gratings with increased efficiency, and (4) larger and more sensitive CCDs that
increase throughput and wavelength coverage relative to SDSS-I\@.
With this new configuration, BOSS is obtaining spectra of approximately
1.5 million luminous galaxies out to redshift $z \simeq 0.7$
from mid-2009 to mid-2014.  This sample complements the
SDSS-I luminous red galaxy (LRG) dataset that extends out to redshift \(z \simeq 0.35\).  LRGs are luminous, elliptical, bulge-dominated galaxies, with remarkably uniform spectral energy distributions that are characterized by a strong break at 4000\AA~\protect\citep{Eisenstein.2001AJ....122.2267E,Eisenstein.2003ApJ...585..694E}. The BOSS luminous galaxy sample is composed of a low-redshift (LOZ) sample designed to select LRGs with \(0.2 \lesssim z \lesssim 0.4\) and a constant mass (CMASS) sample designed to select luminous galaxies with stellar masses in the range \(10^{11}M_{\sun} \lesssim M_\star \lesssim 10^{12}M_{\sun}\) with \(0.4 \lesssim z \lesssim 0.7\)~\protect\citep{Eisenstein:2011AJ....142...72E}. \citet{Masters.2011MNRAS.tmp.1417M} performed visual inspection of 240 BOSS targets in the COSMic Origins Survey field and determined that \(\sim 75\%\) of BOSS galaxies are ``early-type'', with the remainder  ``late-type'', and \(\sim 20\%\) of BOSS galaxies are multiple systems. The combined sample will enable a baryon acoustic feature measurement~\protect\citep{Eisenstein.1998ApJ...496..605E,Eisenstein:2005ApJ...633..560E} of the distance--redshift
relation in several redshift bins.

As a consequence of this design and survey strategy, BOSS presents
an opportunity to discover a sample of strong galaxy--galaxy lenses
at significant cosmological look-back time that is of comparable size
and homogeneity to the SLACS sample at lower redshift.
The two samples in combination may allow a unique study
of the evolution of the structure of mass density and its dependence on
other parameters such as stellar mass and stellar velocity dispersion.

A complementary sample of higher-redshift lenses discovered
in the Canada--France--Hawaii Telescope Strong Lens Legacy Survey
\citep[SL2S; ][]{Cabanac:2007A&A...461..813C,Ruff:2011ApJ...727...96R}
suggests evolution in the slope of the mean-density profile to shallower slopes at higher redshifts.
The key complementarity between SL2S and BOSS strong-lens surveys
is that the former is more uniformly sensitive to selecting lenses across
a range of angular scales (and hence mass scales), whereas the latter
guarantees the spectroscopic redshift completeness for both lensing
and lensed galaxies that is necessary for an accurate conversion
from observed angular scales to physical mass scales.

This paper (Paper I) is the first in a series presenting the initial catalog of 45
spectroscopically selected strong galaxy lens candidates
discovered from the first partial year of BOSS data.
We refer to this survey as the BOSS Emission-Line Lens Survey (BELLS).
Of the initial 45 candidates,
44 were successfully observed with the Wide Field Camera (WFC) of the \textsl{HST}
Advanced Camera for Surveys under Program ID GO-12209 (PI: A.~Bolton),
resulting in 25 confirmed strong gravitational
lenses, along with an additional 11 possible lenses with either very faint
candidate lensed features or significant dust features complicating the analysis.   The remaining target, SDSS\,J0212\(+\)0027, is scheduled for \textsl{HST} visit. 
We show that the BELLS and SLACS lenses trace a single evolving massive galaxy
population in terms of stellar mass, thus confirming the suitability of the
combined sample to constrain the structural and dynamical evolution of massive galaxies
down to the present-day universe.

This paper is organized as follows.
The spectroscopic discovery method is described in \sref{discovery}.
The \textsl{HST} imaging data, reduction method, and the sample of 45 candidates are presented in \sref{acsproc},
and the photometric analysis including the B-spline fits used 
for lens-galaxy subtraction and the \citet{deVaucouleurs.1948AnAp...11..247D}
models used for measuring the magnitudes and half-light radii are presented in \sref{photmod}.
We present our strong-lens modeling methods and results in \sref{lensing},
employing singular isothermal ellipsoid (SIE) mass models in combination with both parametric
and pixelized-grid source-plane surface-brightness models.
A summary of results is presented in \sref{conclusions}.
Throughout this paper, we assume a standard general-relativistic cosmology with parameters
$(h, \Omega_{\mathrm{M}}, \Omega_{\Lambda}) = (0.7, 0.3, 0.7)$~\protect\citep{Larson.2011ApJS..192...16L}. \setcounter{footnote}{1} \renewcommand{\thefootnote}{\fnsymbol{footnote}}
\section{Spectroscopic Candidate Selection} \label{section:discovery} \begin{deluxetable}{lccc}
    \tablecaption{\label{table:emline}Emission-Line Wavelengths}
    \tablehead{
        \colhead{Emission} & \colhead{Restframe} & \colhead{SDSS-I} & \colhead{SDSS-III}\\
        \colhead{Line} & \colhead{Wavelength (\AA)} & \colhead{\(z_{\mathrm{max}}\)} & \colhead{\(z_{\mathrm{max}}\)}\\
        \colhead{\scriptsize (1)} & \colhead{\scriptsize (2)} & \colhead{\scriptsize (3)} & \colhead{\scriptsize (4)}
    }
    \tablecolumns{4}
    \startdata \multirow{2}{*}{[O\,\textsc{ii}]\(\lambda\lambda\)\,3727} & \(3727.09\) & \multirow{2}{*}{\(1.44\)} & \multirow{2}{*}{\(1.63\)}\\
& \(3729.88\) &  & \\
H${\delta}$ & \(4102.89\) & \(1.22\) & \(1.39\)\\
H${\gamma}$ & \(4341.68\) & \(1.10\) & \(1.26\)\\
H${\beta}$ & \(4862.68\) & \(0.87\) & \(1.02\)\\
$$[O\,{\sc iii}]\,4959 & \(4960.30\) & \(0.83\) & \(0.98\)\\
$$[O\,{\sc iii}]\,5007 & \(5008.24\) & \(0.82\) & \(0.96\)\\
$$[N\,{\sc ii}]\,6548 & \(6549.86\) & \(0.39\) & \(0.50\)\\
H${\alpha}$ & \(6564.61\) & \(0.39\) & \(0.49\)\\
$$[N\,{\sc ii}]\,6583 & \(6585.27\) & \(0.38\) & \(0.49\)\\
$$[S\,{\sc ii}]\,6716 & \(6718.29\) & \(0.35\) & \(0.46\)\\
$$[S\,{\sc ii}]\,6730 & \(6732.68\) & \(0.35\) & \(0.46\)
 \enddata
    \tablecomments{Column 1 lists the emission-lines used in the spectroscopic discovery of background emission-line sources and Column 2 shows the rest-frame vacuum wavelength of the emission-line.  Columns 3 and 4 provide the maximum redshift of the emission-line detectable with the SDSS-I and SDSS-III spectrographs, respectively.}
\end{deluxetable}

The premise behind the spectroscopic strong-lens candidate selection of \citet{Bolton:2004AJ....127.1860B} was to search for multiple background emission-lines within the \(3\arcsec\) diameter solid angle covered by the SDSS-I spectroscopic fiber in the residual spectra found after subtracting best-fit galaxy templates to the foreground-galaxy spectrum~\protect\citep[also see][]{Warren:1996MNRAS.278..139W,Hewett:2000ASPC..195...94H,Willis:2005MNRAS.363.1369W,Willis:2006MNRAS.369.1521W}.  For our current search within the BOSS database, we apply much the same method, which is also described in \citet{Bolton:2006ApJ...638..703B,Bolton:2008ApJ...682..964B}, to the \(2\arcsec\) diameter solid angle covered by the SDSS-III spectroscopic fiber.  As a result of the blurring effect of atmospheric seeing, spectroscopic lens selection is efficient even at Einstein radii of \(\theta_E \gtrsim 2\arcsec\)~\protect\citep{Arneson:2011}.

The full list of possible background emission-lines for which we search is given in \tref{emline}, along with the corresponding maximum redshift to which they can be detected with either the SDSS-I or the more red-sensitive BOSS spectrographs.  The spectra are based on SDSS-III imaging~\protect\citep{Gunn.2006AJ....131.2332G,Gunn.1998AJ....116.3040G},  available in the eighth SDSS public Data Release (DR8)~\protect\citep{SDSS3:2011ApJS..193...29A}.

\begin{deluxetable*}{lcccccccc}
    \tabletypesize{\scriptsize}
    \tablewidth{\hsize}
    \tablecaption{\label{table:systems}BOSS Properties of BELLS Candidate Systems}
    \tablehead{
        \colhead{System Name} &
        \colhead{Plate-MJD-Fiber} & 
        \colhead{\(z_{\mathrm{L}}\)} & 
        \colhead{\(z_{\mathrm{S}}\)} & 
        \colhead{{\em i-band} Magnitude} & 
        \colhead{\(R_{\mathrm{eff}}$ (\(\arcsec\))} & 
        \colhead{\(\sigma_{\mathrm{BOSS}}\) (km\,s\(^{-1}\))} & 
        \colhead{Sample} \\
        \colhead{\scriptsize (1)}&\colhead{\scriptsize (2)}&\colhead{\scriptsize (3)}&\colhead{\scriptsize (4)}&\colhead{\scriptsize (5)}&\colhead{\scriptsize (6)}&\colhead{\scriptsize (7)}&\colhead{\scriptsize (8)}
    }
    \tablecolumns{8}
    \startdata SDSS\,J015107.37\(+\)004909.0 & 3606-55182-0679 & \(0.5171\) & \(1.3636\) & \(19.81 \pm 0.05\) & \(0.89 \pm 0.21\) & \(219 \pm 39\) & {\sc cmass} \\ 
SDSS\,J021214.80\(+\)002719.1 & 4236-55479-0603 & \(0.5372\) & \(1.0122\) & \(19.47 \pm 0.03\) & \(1.17 \pm 0.19\) & \(189 \pm 39\) & {\sc cmass} \\ 
SDSS\,J074724.12\(+\)505537.5 & 3677-55205-0551 & \(0.4384\) & \(0.8983\) & \(18.92 \pm 0.03\) & \(1.24 \pm 0.22\) & \(328 \pm 60\) & {\sc cmass} \\ 
SDSS\,J074734.75\(+\)444859.3 & 3676-55186-0581 & \(0.4366\) & \(0.8966\) & \(18.84 \pm 0.03\) & \(2.87 \pm 0.50\) & \(281 \pm 52\) & {\sc cmass} \\ 
SDSS\,J075754.10\(+\)431353.8 & 3676-55186-0063 & \(0.5146\) & \(1.2165\) & \(19.13 \pm 0.04\) & \(2.51 \pm 0.55\) & \ldots & {\sc cmass} \\ 
SDSS\,J080105.30\(+\)472749.6 & 3684-55246-0301 & \(0.4831\) & \(1.5181\) & \(19.87 \pm 0.05\) & \(0.57 \pm 0.17\) & \(\phantom{1}98 \pm 24\) & {\sc cmass} \\ 
SDSS\,J082130.66\(+\)373330.7 & 3760-55268-0699 & \(0.5056\) & \(0.9705\) & \(19.24 \pm 0.02\) & \(0.93 \pm 0.12\) & \(203 \pm 23\) & {\sc cmass} \\ 
SDSS\,J083049.73\(+\)511631.8 & 3695-55212-0142 & \(0.5301\) & \(1.3317\) & \(19.50 \pm 0.04\) & \(1.10 \pm 0.22\) & \(268 \pm 36\) & {\sc cmass} \\ 
SDSS\,J083727.02\(+\)493703.7 & 3697-55290-0255 & \(0.5513\) & \(1.1896\) & \(19.50 \pm 0.03\) & \(1.24 \pm 0.27\) & \(208 \pm 34\) & {\sc cmass} \\ 
SDSS\,J084054.19\(+\)505153.1 & 3697-55290-0921 & \(0.5534\) & \(1.4348\) & \(19.81 \pm 0.03\) & \(0.65 \pm 0.17\) & \(206 \pm 32\) & {\sc cmass} \\ 
SDSS\,J084121.74\(+\)501720.4 & 3697-55290-0087 & \(0.5542\) & \(1.2881\) & \(19.50 \pm 0.03\) & \(0.79 \pm 0.17\) & \(176 \pm 25\) & {\sc cmass} \\ 
SDSS\,J091516.60\(-\)005500.6 & 3766-55213-0675 & \(0.4022\) & \(1.1705\) & \(18.46 \pm 0.02\) & \(1.98 \pm 0.28\) & \(221 \pm 16\) & {\sc cmass} \\ 
SDSS\,J094102.69\(-\)010402.5 & 3782-55244-0513 & \(0.4610\) & \(0.9115\) & \(19.79 \pm 0.04\) & \(0.94 \pm 0.25\) & \(180 \pm 50\) & {\sc cmass} \\ 
SDSS\,J094427.47\(-\)014742.4 & 3782-55244-0268 & \(0.5390\) & \(1.1785\) & \(19.70 \pm 0.04\) & \(1.35 \pm 0.46\) & \(204 \pm 34\) & {\sc cmass} \\ 
SDSS\,J101658.29\(-\)020833.3 & 3784-55269-0237 & \(0.4699\) & \(1.0034\) & \(19.34 \pm 0.04\) & \(1.71 \pm 0.47\) & \(179 \pm 26\) & {\sc cmass} \\ 
SDSS\,J103941.16\(-\)001424.5 & 3833-55290-0109 & \(0.3849\) & \(0.9141\) & \(19.72 \pm 0.04\) & \(0.71 \pm 0.17\) & \(107 \pm 21\) & {\sc cmass} \\ 
SDSS\,J111737.72\(-\)013308.9 & 3788-55246-0930 & \(0.4670\) & \(1.2860\) & \(19.39 \pm 0.07\) & \(2.54 \pm 1.24\) & \(221 \pm 59\) & {\sc cmass} \\ 
SDSS\,J115944.63\(-\)000728.2 & 3843-55278-0069 & \(\phantom{\dagger}0.5793\dagger\) & \(1.3457\) & \(19.63 \pm 0.03\) & \(0.99 \pm 0.22\) & \(165 \pm 41\) & {\sc cmass} \\ 
SDSS\,J121504.44\(+\)004726.0 & 3846-55327-0559 & \(0.6423\) & \(1.2970\) & \(19.27 \pm 0.03\) & \(1.42 \pm 0.21\) & \(262 \pm 45\) & {\sc cmass} \\ 
SDSS\,J122113.28\(-\)022037.8 & 3777-55210-0075 & \(0.4061\) & \(1.0108\) & \(19.07 \pm 0.02\) & \(0.64 \pm 0.09\) & \(240 \pm 24\) & {\sc cmass} \\ 
SDSS\,J122151.92\(+\)380610.5 & 3965-55302-0586 & \(0.5348\) & \(1.2844\) & \(19.77 \pm 0.04\) & \(0.93 \pm 0.22\) & \(187 \pm 48\) & {\sc cmass} \\ 
SDSS\,J123427.99\(-\)024129.6 & 3778-55213-0375 & \(0.4900\) & \(1.0159\) & \(19.18 \pm 0.04\) & \(1.61 \pm 0.46\) & \(122 \pm 31\) & {\sc cmass} \\ 
SDSS\,J131829.39\(-\)010421.6 & 4004-55321-0121 & \(0.6591\) & \(1.3959\) & \(19.66 \pm 0.04\) & \(1.06 \pm 0.31\) & \(177 \pm 27\) & {\sc cmass} \\ 
SDSS\,J133751.31\(+\)362018.1 & 3986-55329-0713 & \(0.5643\) & \(1.1821\) & \(19.01 \pm 0.03\) & \(1.60 \pm 0.33\) & \(225 \pm 35\) & {\sc cmass} \\ 
SDSS\,J134427.96\(+\)325824.6 & 3856-55269-0309 & \(0.4750\) & \(0.9777\) & \(19.57 \pm 0.03\) & \(0.78 \pm 0.13\) & \(198 \pm 22\) & {\sc cmass} \\ 
SDSS\,J134507.63\(-\)012939.8 & 4044-55359-0603 & \(0.4884\) & \(1.2992\) & \(19.92 \pm 0.04\) & \(0.57 \pm 0.13\) & \(200 \pm 25\) & {\sc cmass} \\ 
SDSS\,J134910.30\(+\)361239.7 & 3852-55243-0327 & \(0.4396\) & \(0.8926\) & \(18.88 \pm 0.02\) & \(2.03 \pm 0.31\) & \(178 \pm 18\) & {\sc cmass} \\ 
SDSS\,J135218.99\(+\)321651.8 & 3861-55274-0393 & \(0.4634\) & \(1.0341\) & \(19.28 \pm 0.03\) & \(1.35 \pm 0.33\) & \(161 \pm 21\) & {\sc cmass} \\ 
SDSS\,J145259.14\(+\)332349.4 & 3870-55273-0595 & \(0.5602\) & \(1.1896\) & \(19.19 \pm 0.04\) & \(1.53 \pm 0.31\) & \(296 \pm 42\) & {\sc cmass} \\ 
SDSS\,J150345.69\(+\)322542.1 & 3876-55245-0585 & \(0.6059\) & \(1.4489\) & \(19.49 \pm 0.06\) & \(3.65 \pm 1.18\) & \(162 \pm 68\) & {\sc cmass} \\ 
SDSS\,J152209.54\(+\)291021.9 & 3879-55244-0299 & \(0.5553\) & \(1.3108\) & \(19.67 \pm 0.03\) & \(1.08 \pm 0.27\) & \(166 \pm 27\) & {\sc cmass} \\ 
SDSS\,J153730.27\(+\)022036.9 & 4054-55358-0809 & \(0.4825\) & \(1.0729\) & \(19.62 \pm 0.04\) & \(0.57 \pm 0.15\) & \(199 \pm 25\) & {\sc cmass} \\ 
SDSS\,J154118.56\(+\)181235.1 & 3937-55352-0361 & \(0.5603\) & \(1.1133\) & \(19.82 \pm 0.03\) & \(0.59 \pm 0.00\) & \(174 \pm 24\) & {\sc cmass} \\ 
SDSS\,J154246.33\(+\)162951.8 & 3932-55337-0331 & \(0.3521\) & \(1.0233\) & \(18.39 \pm 0.02\) & \(1.45 \pm 0.13\) & \(210 \pm 16\) & {\sc loz} \\ 
SDSS\,J154503.57\(+\)274805.3 & 3952-55330-0927 & \(0.5218\) & \(1.2886\) & \(18.90 \pm 0.03\) & \(2.65 \pm 0.47\) & \(250 \pm 37\) & {\sc cmass} \\ 
SDSS\,J160113.27\(+\)213833.9 & 3935-55326-0093 & \(0.5435\) & \(1.4461\) & \(19.75 \pm 0.03\) & \(0.63 \pm 0.12\) & \(207 \pm 36\) & {\sc cmass} \\ 
SDSS\,J161109.80\(+\)170526.6 & 4072-55362-0847 & \(0.4766\) & \(1.2109\) & \(19.66 \pm 0.04\) & \(1.33 \pm 0.38\) & \(109 \pm 23\) & {\sc cmass} \\ 
SDSS\,J161523.55\(+\)205636.3 & 4057-55357-0610 & \(0.5894\) & \(1.3103\) & \(18.42 \pm 0.03\) & \(5.92 \pm 1.03\) & \(210 \pm 34\) & {\sc cmass} \\ 
SDSS\,J161846.74\(+\)193026.7 & 4057-55357-0201 & \(0.4906\) & \(1.0164\) & \(19.93 \pm 0.04\) & \(0.76 \pm 0.22\) & \(275 \pm 53\) & {\sc cmass} \\ 
SDSS\,J162754.57\(+\)191004.9 & 4060-55359-0075 & \(0.5251\) & \(1.4892\) & \(19.59 \pm 0.03\) & \(0.59 \pm 0.11\) & \(236 \pm 33\) & {\sc cmass} \\ 
SDSS\,J163150.33\(+\)185404.1 & 4064-55366-0790 & \(0.4081\) & \(1.0863\) & \(17.70 \pm 0.01\) & \(2.07 \pm 0.10\) & \(272 \pm 14\) & {\sc loz} \\ 
SDSS\,J163714.58\(+\)143930.1 & 4065-55368-0515 & \(0.3910\) & \(0.8744\) & \(19.49 \pm 0.03\) & \(0.89 \pm 0.15\) & \(208 \pm 30\) & {\sc cmass} \\ 
SDSS\,J212252.04\(+\)040935.5 & 4081-55365-0241 & \(0.6261\) & \(1.4517\) & \(19.29 \pm 0.03\) & \(1.76 \pm 0.25\) & \(324 \pm 56\) & {\sc cmass} \\ 
SDSS\,J212510.67\(+\)041131.6 & 4080-55471-0621 & \(0.3632\) & \(0.9777\) & \(18.21 \pm 0.01\) & \(1.47 \pm 0.09\) & \(247 \pm 17\) & {\sc loz} \\ 
SDSS\,J230335.17\(+\)003703.2 & 4207-55475-0933 & \(0.4582\) & \(0.9363\) & \(18.96 \pm 0.02\) & \(1.35 \pm 0.23\) & \(274 \pm 31\) & {\sc cmass} \\ 
 \enddata
    \tablecomments{Column 1 provides the SDSS System Name in terms of truncated J2000 R.A. and decl. in the
    format HHMMSS.ss\(\pm\)DDMMSS.s.  Column 2 provides the Plate-MJD-Fiber of the primary spectrum for the BOSS target~\protect\citep{Eisenstein:2011AJ....142...72E} as of v5\_4\_45 of the BOSS pipeline.  Columns 3 and 4 provide the foreground BOSS spectroscopic redshift and the spectroscopically detected background emission-line source redshift, respectively. Columns 5 and 6 provide the BOSS de~Vaucouleurs model \(i\)-band magnitudes and intermediate axis effective radii, respectively~\protect\citep{Fukugita.1996AJ....111.1748F,SDSS3:2011ApJS..193...29A}.  Column 7 provides the velocity dispersion computed by the BOSS pipeline (v5\_4\_45), uncorrected for aperture effects, using redshift-error marginalization and restricted stellar template sets as described in \citet{Shu.2011arXiv1109.6678S}.  Column 8 indicates whether the candidate was selected from the BOSS LOZ or CMASS sample.  All the BELLS candidates in this table were observed (or will be observed in the case of SDSS\,J0212\(+\)0027) by {\em HST} cycle 18 Program ID GO-12209.  The \textsl{HST}-ACS \(I_{814}\)-band properties are listed in \tref{photmod}, including our classification of morphology, multiplicity, and lens grade, with our justification of BELLS confirmed gravitational lenses in \tref{lensing:grade}. \(\dagger\)~The lens redshift quoted for SDSS\,J1159\(-\)0007 differs from the BOSS v5\_4\_45 galaxy redshift that will be reported in the ninth SDSS public Data Release (DR9) as \(z\_{noqso}\), due to the background emission-line flux.}
\end{deluxetable*}

\figuretitle{Spectroscopically discovered background emission-lines in BELLS target galaxies}
\begin{figure*}[t]
    \plotone{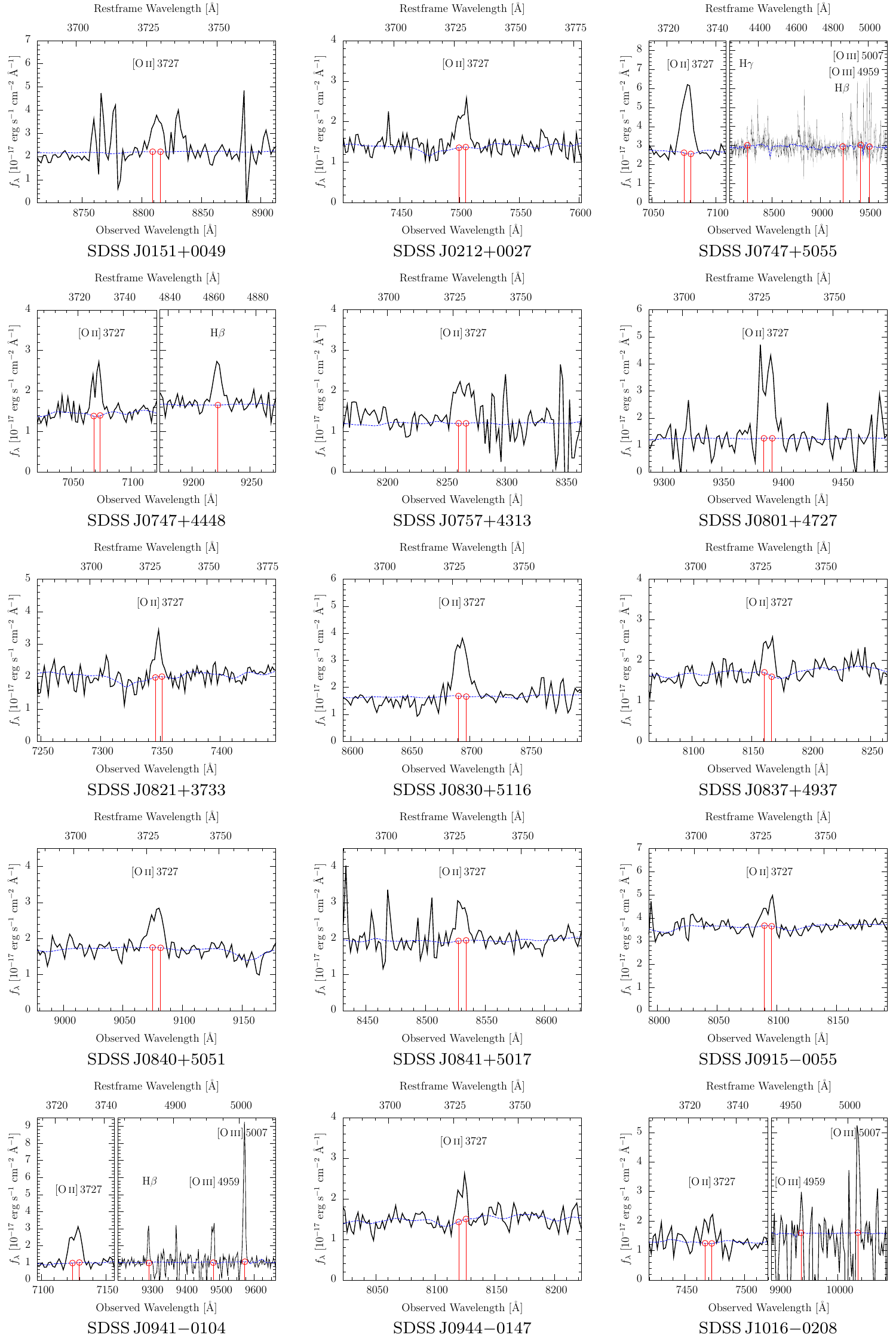} 
    \caption{\label{figure:emlinehit}\thefiguretitle. The black solid-line shows the BOSS observed flux density, \(f_\lambda\), as a function of observed (lower axis) and rest-frame (upper axis) wavelengths.  The blue dashed-line shows the BOSS template fitted to the continuum of the foreground-galaxy, and the red vertical line show the wavelength of the background emission-lines discovered in the differences between the observed flux and the best-fit template. This figure is \emph{continued}.}
\end{figure*} \addtocounter{figure}{-1}
\begin{figure*}[p] \plotone{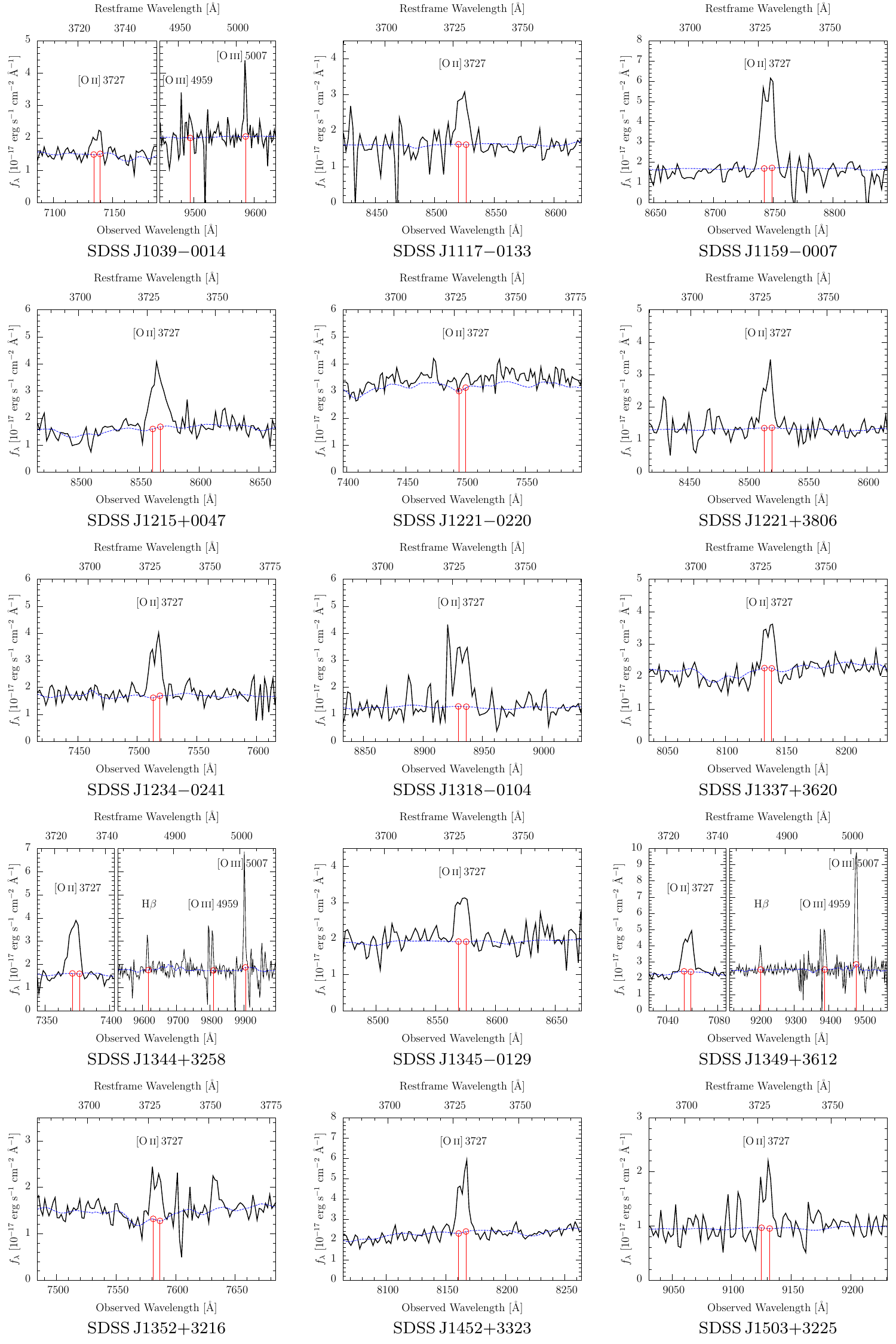} \caption{\emph{Continued}. \thefiguretitle.}\end{figure*} \addtocounter{figure}{-1}
\begin{figure*}[p] \plotone{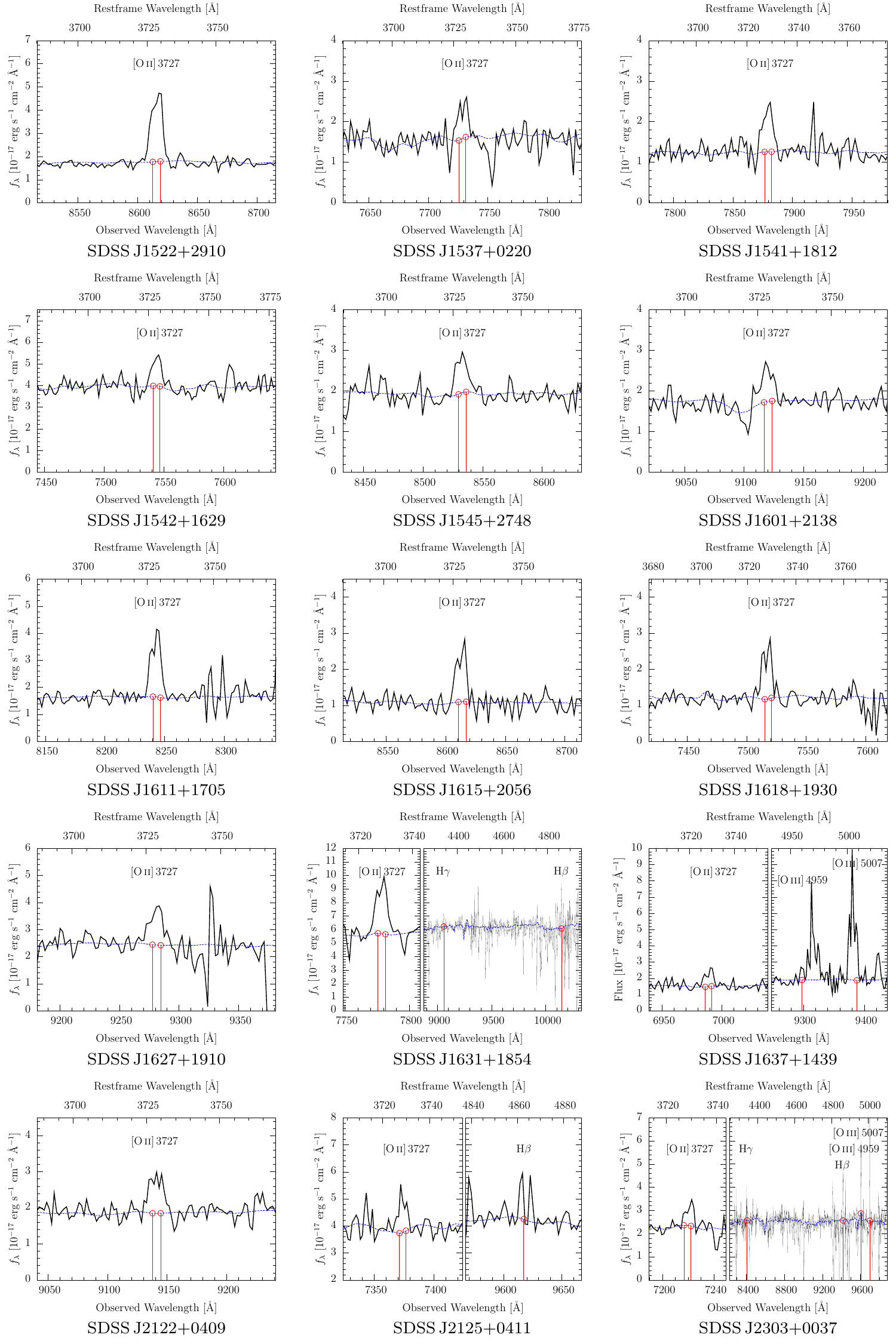} \caption{\emph{Continued}. \thefiguretitle.}\end{figure*}

\figuretitle{BELLS Strong Gravitational Lens Galaxies and Foreground Subtracted Images}
\begin{figure*}[t]
    \plotone{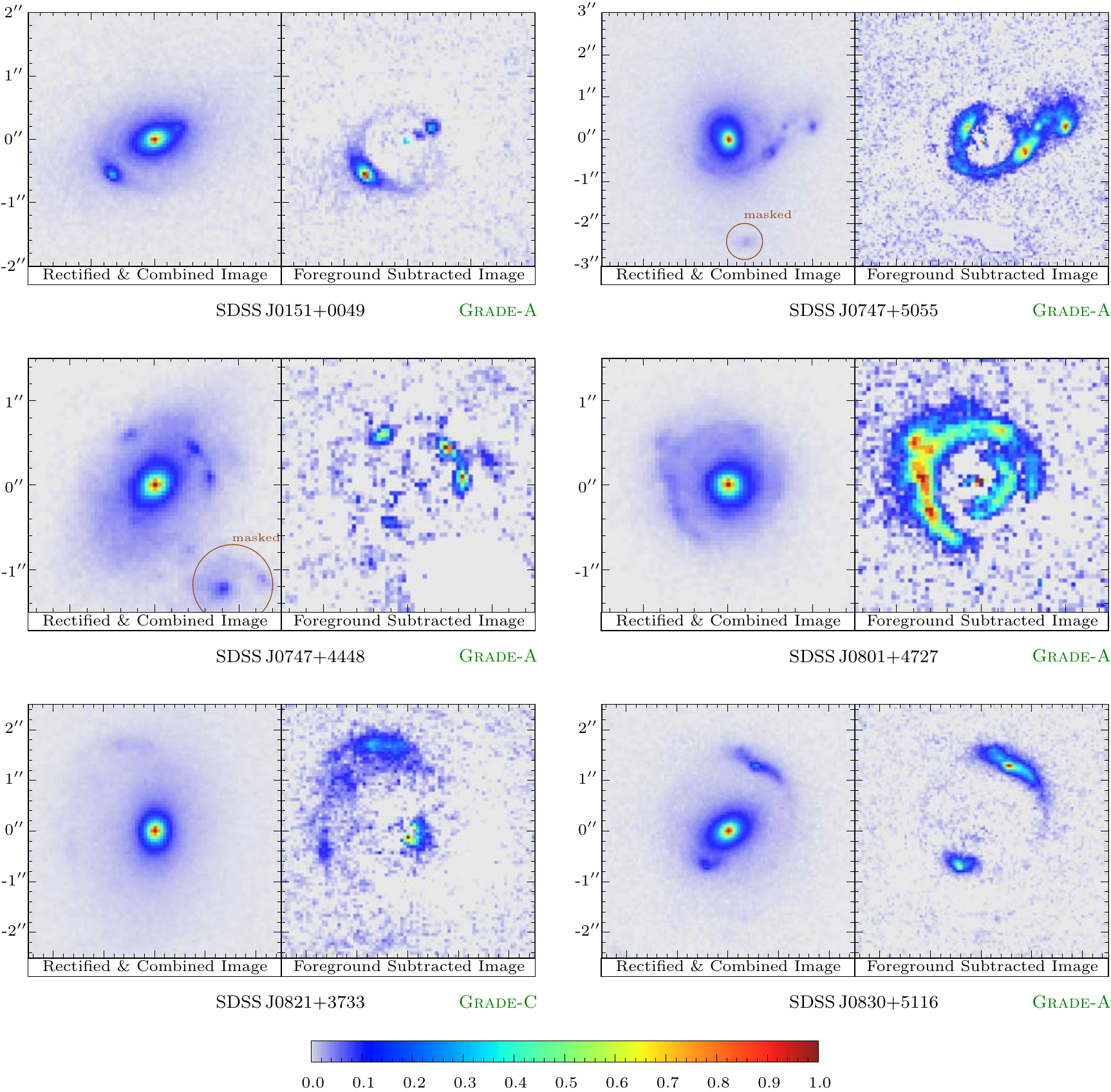} 
    \caption{\label{figure:acsproc}\thefiguretitle. The 36 Grade-\,A, B and C lenses discovered under \textsl{HST} Cycle 18 program 12209.  For each system, the left panel shows the \textsl{HST} ACS-WFC F814W rectified and combined images with north up and east to the left, and the right panel reveals the lensed features which remain after the foreground-galaxy has been subtracted by the B-spline method described in \sref{acsproc}.  Probable non-lensed extraneous features that were masked are circled. System properties from the BOSS data are listed in \tref{systems}, and those from the \textsl{HST}-ACS data are listed in \tref{photmod}. Comments justifying our lens grade are provided in \tref{lensing:grade}. This figure is \emph{continued}.}
\end{figure*} \addtocounter{figure}{-1}
\begin{figure*}[p] \plotone{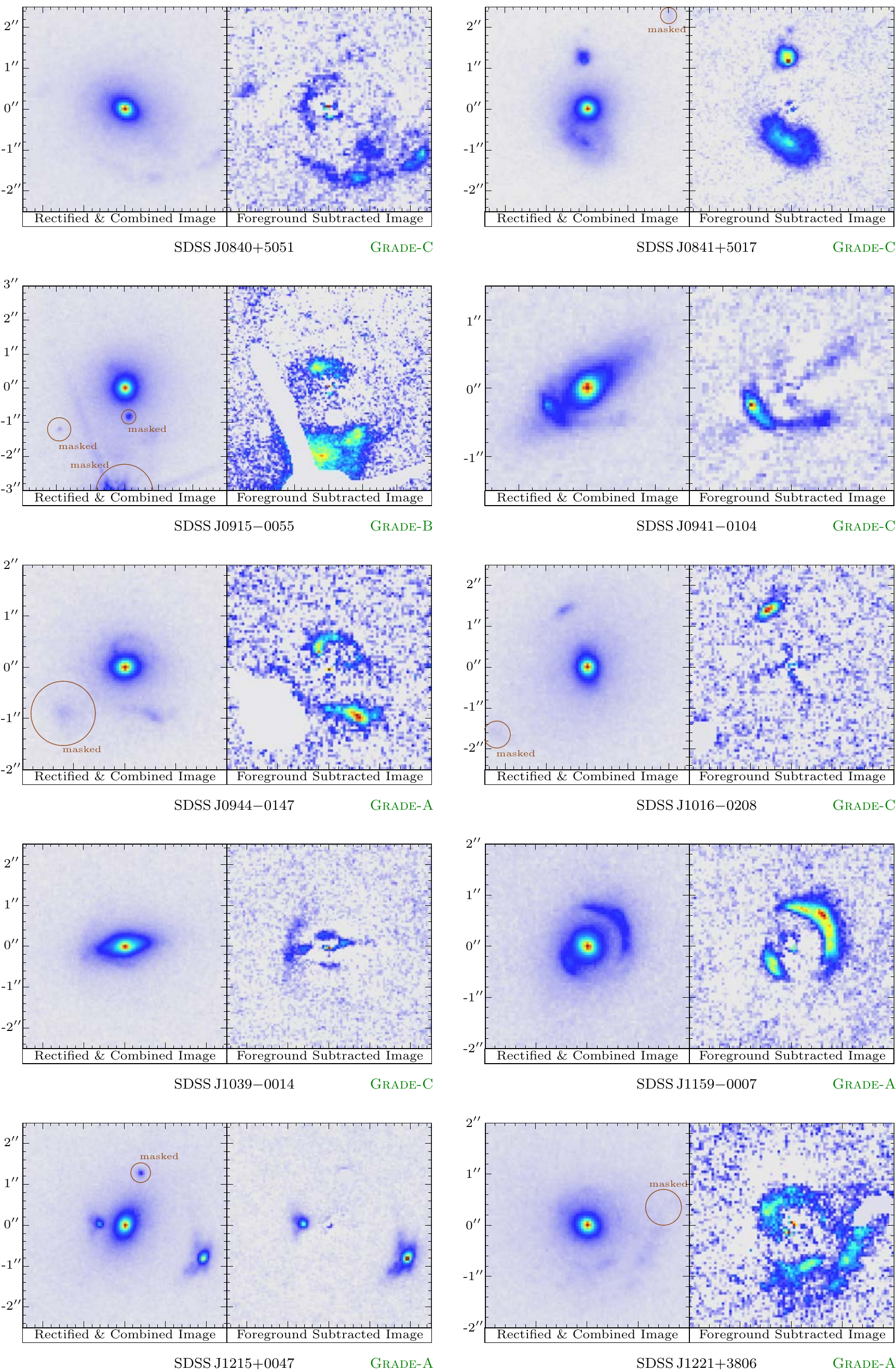} \caption{\emph{Continued}. \thefiguretitle.}\end{figure*} \addtocounter{figure}{-1}
\begin{figure*}[p] \plotone{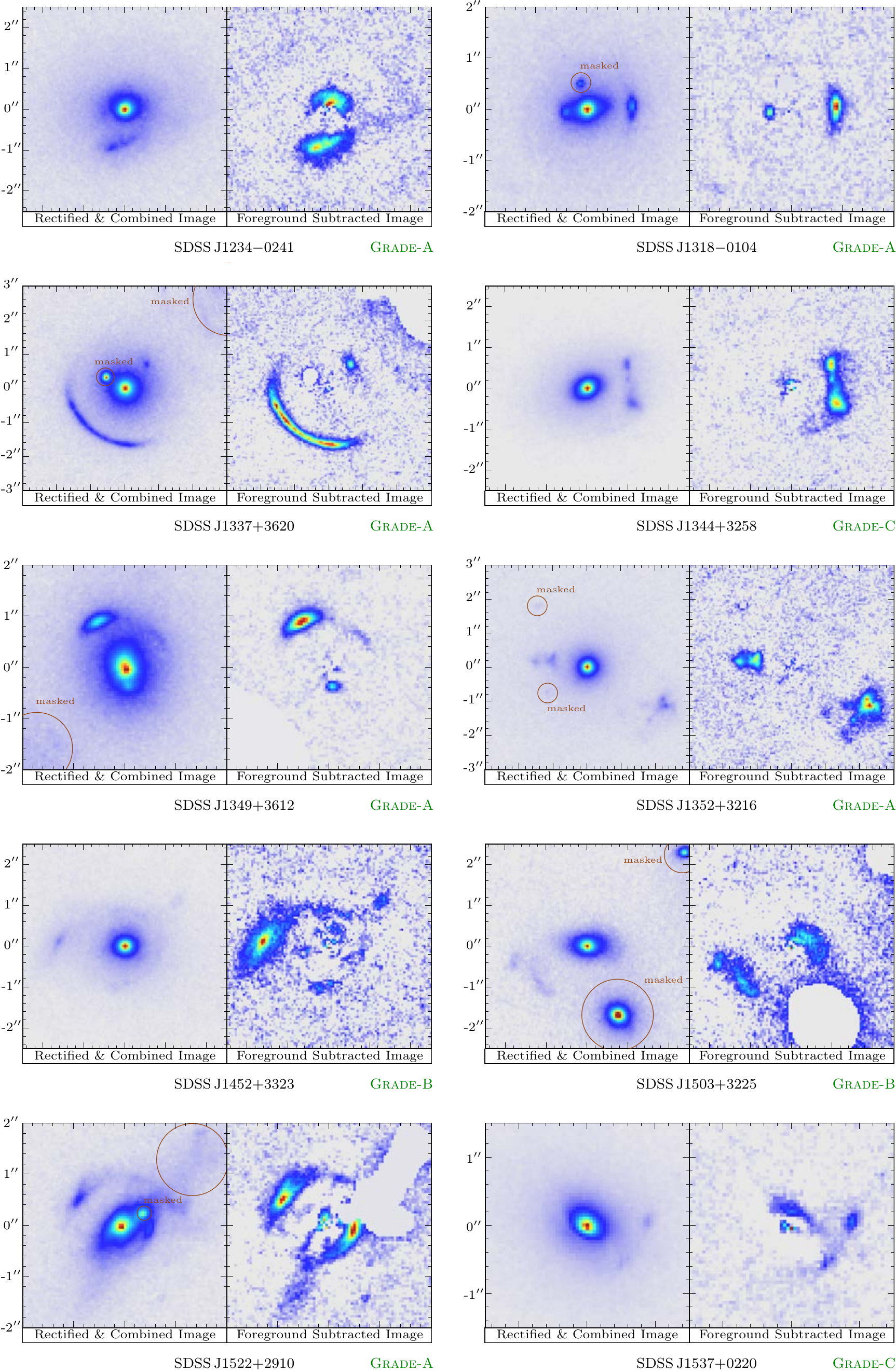} \caption{\emph{Continued}. \thefiguretitle.}\end{figure*} \addtocounter{figure}{-1}
\begin{figure*}[p] \plotone{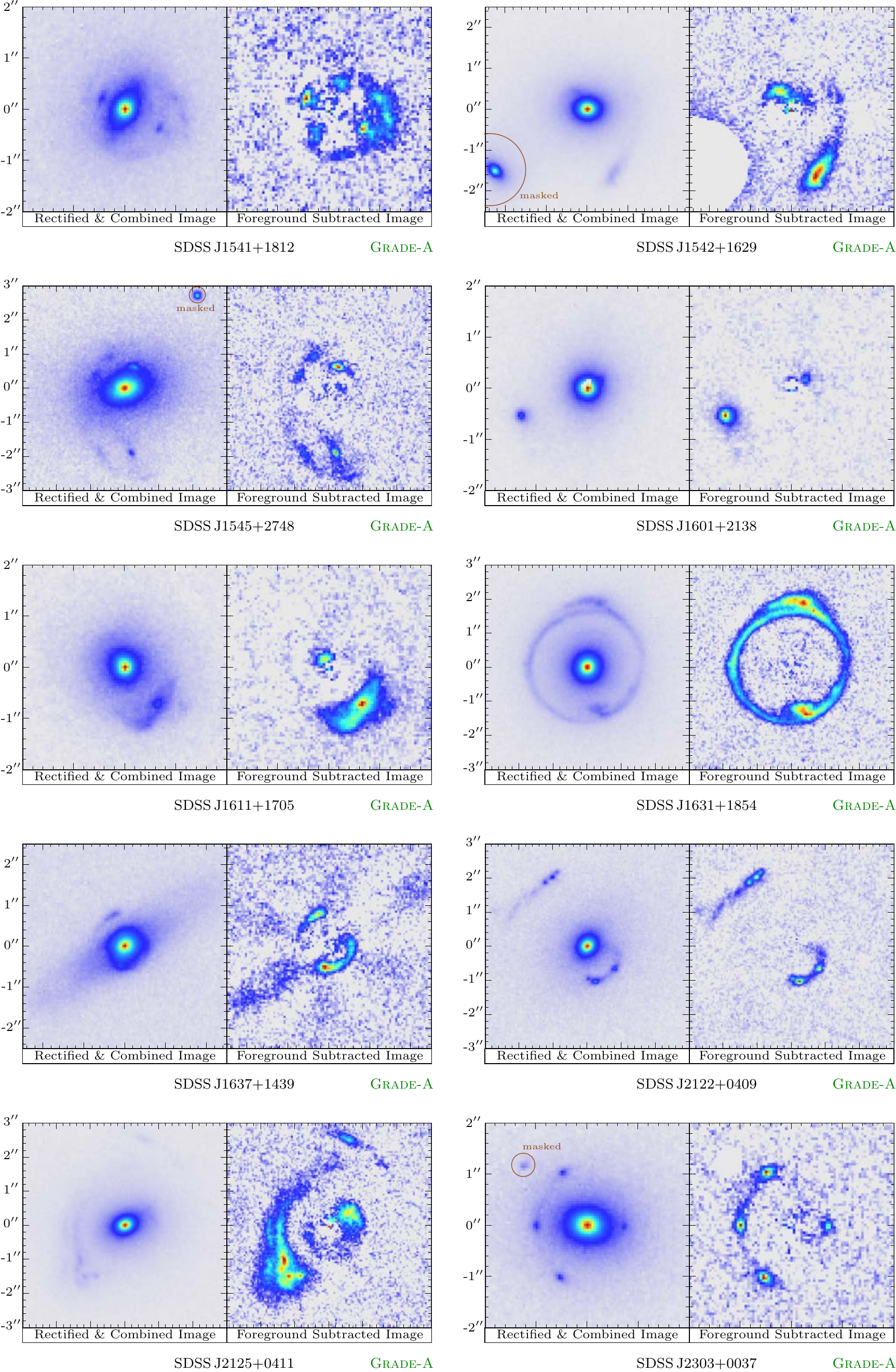} \caption{\emph{Continued}. \thefiguretitle.}\end{figure*}

We summarize our selection procedure here for completeness, and to reflect minor changes in the implementation relevant for BOSS.

\begin{enumerate*}
\item Rescale the spectrum noise-vector estimates based upon the residual fluctuations in sky-subtracted sky-fiber spectra on the plate.

\item Select all objects both targeted and spectroscopically classified as galaxies, without any redshift warning flags.  In the redshift analysis, only ``galaxy'' and ``star'' templates were allowed, since the inclusion of ``qso'' templates in the analysis of relatively low signal-to-noise ratio (S/R) BOSS galaxy spectra yields an unacceptable number of spurious classifications and redshifts.  There were 133,852 galaxy spectra scanned in total.

\item Mask the positions of common emission-lines in the spectra of the target-galaxies, re-fit a model to the BOSS target galaxy continuum using a basis of seven principal component analysis (PCA) eigenspectra generated as described in \citet{SDSS3:2011ApJS..193...29A} (versus four PCA eigenspectra used in the redshift analysis), and subtract this continuum model from the data.

\item \label{item:discovery:linescan} Scan the residual spectra automatically for at least two of the lines H$\alpha$, H$\beta$, [O\,\textsc{iii}]\,5007, [O\,\textsc{iii}]\,4959, or [O\,\textsc{ii}]\(\lambda\lambda\)\,3727 detected at \(4\sigma\) significance or higher at a common redshift greater than that of the BOSS target.  We refer to these as ``multi-line'' detections. There were 2898 multi-line hits.

\item Scan the residual spectra automatically for a single line detected at \(6\sigma\) significance or higher that would be at a higher redshift than the BOSS target if identified as [O\,\textsc{ii}]\(\lambda\lambda\)\,3727.  We refer to these as ``single-line'' detections.  There were 8211 single-line hits.

\item \label{item:discovery:prune.sky} Bin the single-line detections by observed-frame wavelength, and prune those that are in highly overpopulated bins associated with night sky emission-line subtraction residuals.

\item \label{item:discovery:prune.fmask} Bin the single-line detections by target-galaxy rest-frame wavelength, and prune those that are in highly overpopulated bins associated with imperfect foreground-galaxy feature masking and subtraction.  There were 2137 single-line hits remaining after the cuts of \irefs{discovery:prune.sky}{discovery:prune.fmask}.

\item Veto any single-line [O\,\textsc{ii}]\(\lambda\lambda\)\,3727 identification that is more plausibly explained as H$\alpha$, H$\beta$, or [O\,\textsc{iii}]\,5007, based upon the appearance of secondary lines at \(3\sigma\) significance or higher.  There were 1624 single-line hits remaining.

\item Remove any single-line detections also found in the multi-line search.  There were 1268 single-line hits remaining.  

\item Remove any detections in which the background redshift is not greater than the foreground redshift by the margin, \(z_{\mathrm{L}} - z_{\mathrm{S}} \ge 0.05\).  There were 1303 multi-line hits and 741 single-line hits remaining.

\item Visually inspect the reduced spectra of all remaining selected systems at the position of the candidate background emission-lines, to remove obvious data-reduction/data-quality artifacts.   There were 1226 multi-line hits and 507 single-line hits remaining after this step.

\item Inspect neighboring fiber spectra on each plate at the wavelength of the candidate background emission to rule out cross talk from bright emission-lines in neighboring spectra, and spatially localized auroral emission not subtracted by the global sky spectrum model.

\item Inspect the individual exposure spectra that were co-added to give the final detection spectrum, to ensure that the candidate background emission-line feature is not present in only a single frame (as would be expected for e.g., cosmic-ray hits).

\item Inspect the raw spectroscopic CCD data for all exposures at the position of [O\,\textsc{ii}]\(\lambda\lambda\)\,3727
emission for single-line candidates, to ensure that the detection is not related to cosmetic blemishes in the CCD.

\item Compute a total S/R for all detections as the quadrature sum of the individual S/R values (i.e., \(\sigma\)-values) of all lines in \tref{emline} detected at a significance of \(3\sigma\) or higher --- including those which we do not employ as primary detection triggers in \iref{discovery:linescan} --- and rank candidates by this total S/R.
\end{enumerate*}

In addition to ranking candidates by total S/R, we also select candidates to maximize lensing probability.  \citet{Bolton:2008ApJ...682..964B} show empirically that the lensing probability of a candidate system is a strongly increasing function of the predicted Einstein radius of the system using a simple singular isothermal sphere model based on the foreground-galaxy velocity dispersion and the redshifts of both foreground- and background-galaxies.

For BOSS lens candidates, we do not have precise velocity dispersion measurements to use in this calculation nor do we want to introduce a selection bias by using luminosity as a proxy.  Hence, we assume a fiducial velocity dispersion of 250 km\,s$^{-1}$ for all candidates and calculate a predicted ``Einstein radius'' given the foreground and background redshifts.  This fiducial value is taken from the median of all well-measured velocity dispersions in the subset of the \citet{Bolton:2008ApJ...682..964B} SLACS lens sample having early-type morphology.  We then compute the physical radius subtended by this predicted Einstein radius at the redshift of the lens galaxy and require it to be greater than 4.5 kpc.  Since no actual magnitude or velocity dispersion information is used to make this cut, it can only bias the sample in terms of foreground and background redshift distributions.  The cut corresponds to a predicted Einstein radius of \(0\farcs84\) at $z_{\mathrm{L}} = 0.4$, and \(0\farcs63\) at $z_{\mathrm{L}} = 0.7$.  We verified that the sample selected by this cut does produce a reasonably uniform distribution of candidates over the available redshift range in the parent sample. The fiducial values of \(\sigma_v = 250\) km\,s$^{-1}$ and \(\theta_{E}=4.5\) kpc are not especially significant and can be traded off against one another with no effect on the selected sample.  Effectively, our cut serves to favor candidates with higher background redshifts at a given foreground redshift, and pushes to slightly smaller angular Einstein radii for higher-redshift lens galaxies.
 
These steps were carried out for all BOSS survey plates available prior to 2010 June 21: 239 unique plates, with an average of approximately 500 successful galaxy redshifts per plate. The systems are listed in \tref{systems}, including the BOSS pipeline  redshift, the redshift of our candidate background emission-line source, and the BOSS photometry and velocity dispersion measurements.  The background emission-line spectra remaining after subtracting the best-fit foreground spectra are presented in \fref{emlinehit}.  The original detection of SDSS\,J1221\(-\)0220 was an artifact of the spectroscopic data-reduction version used at the time of the \textsl{HST} target selection.

\section{Image Processing} \label{section:acsproc} 
The 45 systems identified in \tref{systems} were submitted for observation under \textsl{HST} Cycle 18 Program ID GO-12209, with one single-orbit visit per target using the $I$-band F814W filter, with four sub-exposures of approximately 500 s each in the ACS-WFC-DITHER-BOX pattern.

After each complete (and successful) visit,  the individual flat-fielded (FLT) sub-exposure files generated by the Space Telescope Science Institute (STScI) {\tt CALACS} pipeline were downloaded, and reduced using {\tt ACSPROC} --- a custom-built pipeline that was generalized from the SLACS version of the software described in \citet{Bolton:2008ApJ...682..964B}.  For each visit, the steps in the current version of {\tt ACSPROC}, implemented in the {\tt IDL} programming language, are as follows.

\begin{enumerate*}
\item Extract a \(1500 \times 1500\) pixel section of the FLT file for each sub-exposure, centered on the target, performing cosmic-ray rejection using {\tt L.A. COSMIC}~\protect\citep{vanDokkum:2001PASP..113.1420V}, and sky-level subtraction using the FLT {\tt MDRIZSKY} {\tt FITS} header, as determined by the STScI pipeline {\tt MULTIDRIZZLE} reduction.
\item Obtain the approximate pixel-shift across sub-exposures through fast Fourier transform cross correlation and compute the centroid of the target galaxy for each sub-exposure by fitting an elliptical Moffat profile using the {\tt MPFIT2DPEAK} nonlinear least-squares fitting function.
\item Rectify the sub-exposures onto a uniform \(0\farcs05\) per pixel grid, with north up and east left, centered on the computed centroid. Rectify, with identical sampling, an appropriate model point-spread function (PSF) which was computed using the {\tt Tiny Tim} \textsl{HST} PSF simulator~\protect\citep{Krist:1993ASPC...52..536K}.
\item \label{item:acsproc:comb} Combine the sub-exposures into a single stacked image, with additional cosmic ray rejection. Similarly, combine the PSF model.
\item \label{item:acsproc:mask} Within a \(280 \times 280\) (\(14\arcsec \times 14\arcsec\)) section of the stacked image centered on the centroid of the target, separately mask potential lensed features, and probable non-lensed stars and galaxies.  
\item Fit a set of elliptical radial B-spline models~\protect\citep{Bolton:2006ApJ...638..703B,Bolton:2008ApJ...682..964B} to the masked stacked image using the {\tt MPFIT} nonlinear least-squares implementation of the Levenberg--Marquardt algorithm~\protect\citep{More:1993}, allowing for either none, or combinations of higher-order multipole moments, including the dipole, quadrupole, and octopole moments.  Each of the B-spline models are manually inspected, and the masks are iteratively refined by the appearance of strong-lensing features in the B-spline-subtracted residuals.  The B-spline fit which leads to the lowest \(\chi^2\) without introducing obviously spurious small-scale structure is selected as the model for lens galaxy subtraction.
\end{enumerate*}

\iref{acsproc:comb} leads to a combined and rectified image for all of the systems listed in \tref{systems}.  \iref{acsproc:mask} leads to a Grade-A foreground-subtracted image for systems with clear and convincing evidence of multiple imaging for a subset of those systems.  Systems with lensed features that are too low in surface brightness or contaminated by significant bright star light are classified as Grade-B systems and systems that are unsuitable for lens modeling due to the absence of clearly identifiable counter-images are classified as Grade-C systems.  The systems without any identifiable lensed features are classified as Grade-X systems for which we do not complete \iref{acsproc:mask}.

The results of the {\tt ACSPROC} image processing, including the combined and rectified stacked images and the foreground-subtracted images are shown in \fref{acsproc}  for all of the Grade-\,A, B and C systems.  The combined and rectified stacked images for the Grade-X systems are included in the Appendix \fref{appendix:GradeX} for completeness.

\begin{figure}[h]
   \plotone{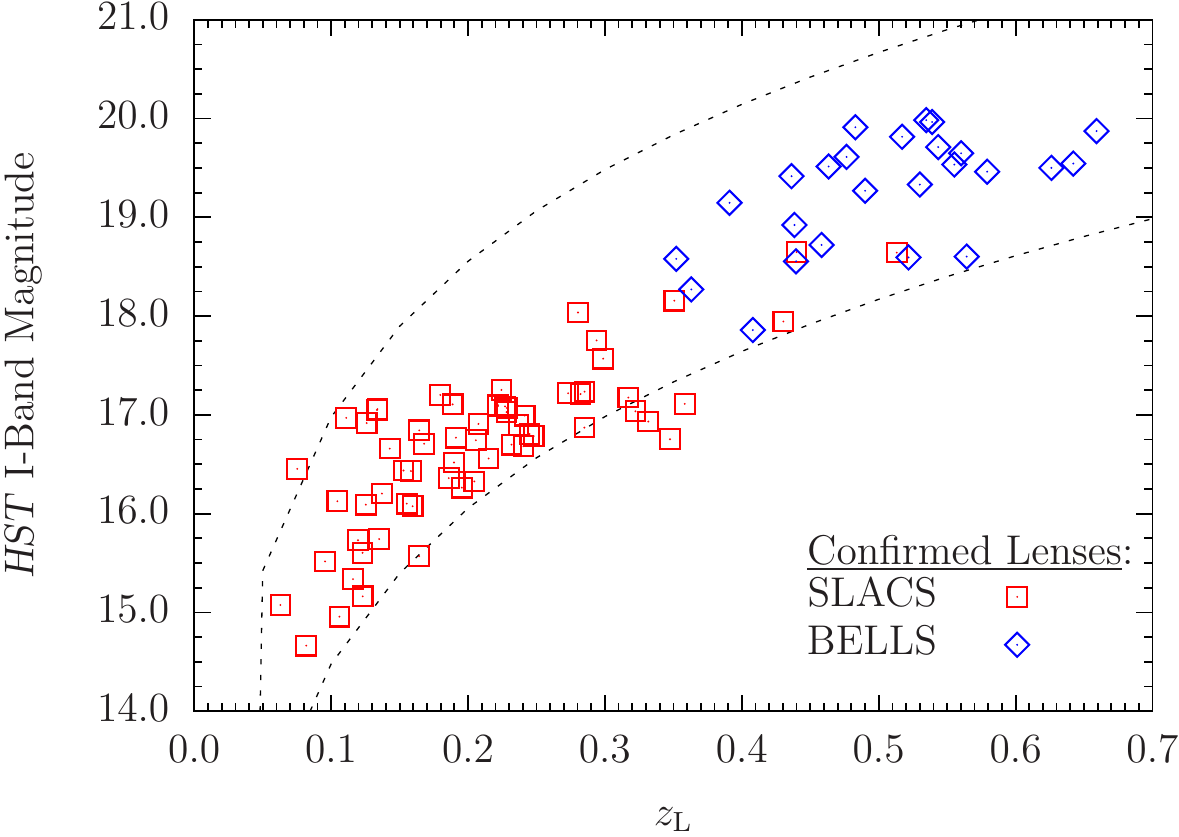}
   \caption{\label{figure:population.magnitude} Distribution of confirmed SLACS and BELLS lens \textsl{HST} \(I_{814}\)-band magnitudes,  shown as a function of lens redshift.  SLACS lenses are shown with red squares, and BELLS lenses are shown with blue diamonds.  Stellar masses are indicated by the \(10^{11} M_{\sun}\) upper track and \(10^{12} M_{\sun}\) lower track with black dashed-lines, based upon the stellar population model of \citet{Maraston:2009MNRAS.394L.107M}.} 
\end{figure}

\begin{figure}[h]
   \plotone{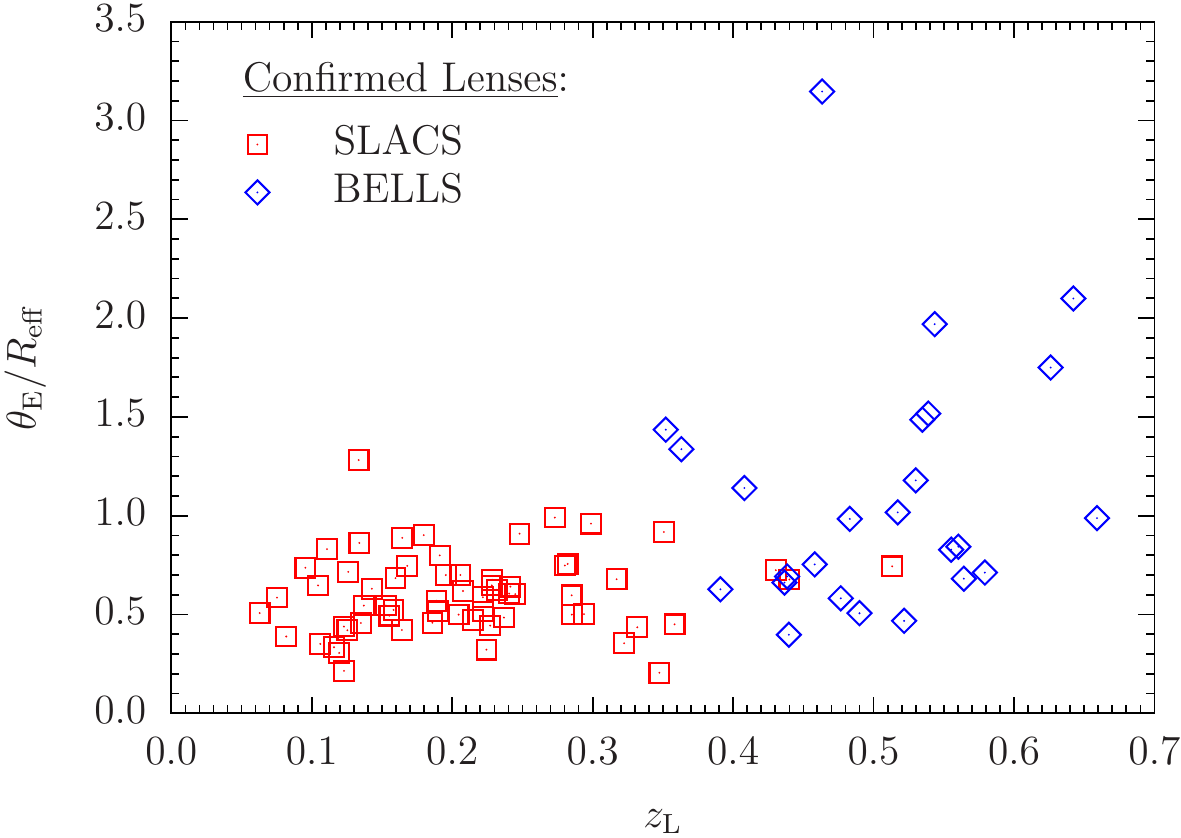}
   \caption{\label{figure:population.size} Distribution of the ratio between the Einstein radii and the effective radii, as a function of lens redshift.  SLACS lenses are shown with red squares, and BELLS lenses are shown with blue diamonds.} 
\end{figure}

\section{Photometric Analysis} \label{section:photmod} \begin{deluxetable*}{ccccccc}
    \tabletypesize{\scriptsize}
    \tablewidth{\hsize}
    \tablecaption{\label{table:photmod}\textsl{HST}-ACS \(I_{814}\)-band Properties of BELLS Systems}
    \tablehead{
        \colhead{System Name} &
        \colhead{\(m_{I_{814}}\) (Obs.)} & 
        \colhead{\(\Delta m_{I_{814}}\) (extin.)} & 
        \colhead{\(R_{\mathrm{eff}}\) (\(\arcsec\))} & 
        \colhead{\(q\)} & 
        \colhead{\(PA\) (\(\degr\))} &
        \colhead{Classification} \\
        \colhead{\scriptsize (1)} & \colhead{\scriptsize (2)} & \colhead{\scriptsize (3)} & \colhead{\scriptsize (4)} & \colhead{\scriptsize (5)} & \colhead{\scriptsize (6)} & \colhead{\scriptsize (7)}
    }
    \tablecolumns{7}
    \startdata SDSS\,J0151\(+\)0049 & \(19.816\) & \(0.049\) & \(0.665 \pm 0.002\) & \(0.604 \pm 0.001\) & \(109.0 \pm 0.1\) & E-S-A \\ 
SDSS\,J0747\(+\)5055 & \(18.923\) & \(0.111\) & \(1.089 \pm 0.002\) & \(0.737 \pm 0.001\) & \(\phantom{00}6.9 \pm 0.1\) & E-S-A \\ 
SDSS\,J0747\(+\)4448 & \(19.417\) & \(0.066\) & \(0.924 \pm 0.002\) & \(0.645 \pm 0.001\) & \(140.2 \pm 0.1\) & E-S-A \\ 
SDSS\,J0757\(+\)4313 & \(18.500\) & \(0.068\) & \(3.818 \pm 0.009\) & \(0.572 \pm 0.001\) & \(\phantom{0}49.8 \pm 0.1\) & L-M-X \\ 
SDSS\,J0801\(+\)4727 & \(19.911\) & \(0.103\) & \(0.499 \pm 0.001\) & \(0.951 \pm 0.002\) & \(\phantom{0}70.8 \pm 1.3\) & E-S-A \\ 
SDSS\,J0821\(+\)3733 & \(19.219\) & \(0.079\) & \(0.551 \pm 0.001\) & \(0.743 \pm 0.001\) & \(172.8 \pm 0.1\) & E-S-C \\ 
SDSS\,J0830\(+\)5116 & \(19.332\) & \(0.087\) & \(0.969 \pm 0.002\) & \(0.699 \pm 0.001\) & \(121.6 \pm 0.1\) & E-S-A \\ 
SDSS\,J0837\(+\)4937 & \(19.462\) & \(0.060\) & \(0.669 \pm 0.001\) & \(0.473 \pm 0.001\) & \(\phantom{0}51.3 \pm 0.1\) & L-S-X \\ 
SDSS\,J0840\(+\)5051 & \(19.984\) & \(0.043\) & \(0.357 \pm 0.001\) & \(0.699 \pm 0.001\) & \(\phantom{0}63.7 \pm 0.1\) & E-S-C \\ 
SDSS\,J0841\(+\)5017 & \(19.624\) & \(0.045\) & \(0.648 \pm 0.001\) & \(0.939 \pm 0.002\) & \(\phantom{0}79.3 \pm 0.8\) & E-S-C \\ 
SDSS\,J0915\(-\)0055 & \(18.518\) & \(0.070\) & \(1.219 \pm 0.002\) & \(0.925 \pm 0.001\) & \(173.2 \pm 0.5\) & E-S-B \\ 
SDSS\,J0941\(-\)0104 & \(20.005\) & \(0.063\) & \(0.458 \pm 0.001\) & \(0.568 \pm 0.001\) & \(129.9 \pm 0.1\) & L-S-C \\ 
SDSS\,J0944\(-\)0147 & \(19.965\) & \(0.067\) & \(0.478 \pm 0.001\) & \(0.785 \pm 0.002\) & \(\phantom{0}91.7 \pm 0.2\) & E-S-A \\ 
SDSS\,J1016\(-\)0208 & \(19.797\) & \(0.072\) & \(0.465 \pm 0.001\) & \(0.816 \pm 0.002\) & \(\phantom{00}1.5 \pm 0.3\) & E-S-C \\ 
SDSS\,J1039\(-\)0014 & \(19.345\) & \(0.094\) & \(0.812 \pm 0.001\) & \(0.440 \pm 0.001\) & \(\phantom{0}97.9 \pm 0.0\) & L-S-C \\ 
SDSS\,J1117\(-\)0133 & \(18.896\) & \(0.079\) & \(2.397 \pm 0.006\) & \(0.938 \pm 0.002\) & \(135.3 \pm 1.0\) & U-U-X \\ 
SDSS\,J1159\(-\)0007 & \(19.463\) & \(0.049\) & \(0.958 \pm 0.002\) & \(0.966 \pm 0.002\) & \(\phantom{0}23.5 \pm 1.9\) & E-S-A \\ 
SDSS\,J1215\(+\)0047 & \(19.544\) & \(0.050\) & \(0.651 \pm 0.001\) & \(0.684 \pm 0.001\) & \(152.9 \pm 0.1\) & E-S-A \\ 
SDSS\,J1221\(-\)0220 & \(18.942\) & \(0.055\) & \(0.710 \pm 0.001\) & \(0.586 \pm 0.001\) & \(\phantom{0}54.9 \pm 0.1\) & E-S-X \\ 
SDSS\,J1221\(+\)3806 & \(19.984\) & \(0.029\) & \(0.470 \pm 0.001\) & \(0.838 \pm 0.002\) & \(\phantom{0}65.6 \pm 0.3\) & E-S-A \\ 
SDSS\,J1234\(-\)0241 & \(19.269\) & \(0.064\) & \(1.054 \pm 0.002\) & \(0.762 \pm 0.002\) & \(\phantom{0}93.7 \pm 0.2\) & E-S-A \\ 
SDSS\,J1318\(-\)0104 & \(19.873\) & \(0.050\) & \(0.687 \pm 0.002\) & \(0.761 \pm 0.002\) & \(106.8 \pm 0.3\) & E-S-A \\ 
SDSS\,J1337\(+\)3620 & \(18.603\) & \(0.023\) & \(2.034 \pm 0.003\) & \(0.960 \pm 0.001\) & \(111.4 \pm 1.0\) & E-S-A \\ 
SDSS\,J1344\(+\)3258 & \(19.581\) & \(0.031\) & \(0.524 \pm 0.001\) & \(0.746 \pm 0.001\) & \(117.8 \pm 0.1\) & E-S-C \\ 
SDSS\,J1345\(-\)0129 & \(21.877\) & \(0.071\) & \(1.000 \pm 0.003\) & \(0.000 \pm 0.001\) & \(\phantom{00}0.0 \pm 0.1\) & E-S-X \\ 
SDSS\,J1349\(+\)3612 & \(18.555\) & \(0.025\) & \(1.886 \pm 0.003\) & \(0.743 \pm 0.001\) & \(\phantom{00}8.7 \pm 0.1\) & E-S-A \\ 
SDSS\,J1352\(+\)3216 & \(19.514\) & \(0.024\) & \(0.579 \pm 0.001\) & \(0.949 \pm 0.001\) & \(127.2 \pm 0.8\) & E-S-A \\ 
SDSS\,J1452\(+\)3323 & \(19.487\) & \(0.026\) & \(0.623 \pm 0.001\) & \(0.836 \pm 0.001\) & \(\phantom{0}87.9 \pm 0.2\) & E-S-B \\ 
SDSS\,J1503\(+\)3225 & \(20.118\) & \(0.032\) & \(0.769 \pm 0.003\) & \(0.625 \pm 0.002\) & \(\phantom{0}82.4 \pm 0.2\) & E-M-B \\ 
SDSS\,J1522\(+\)2910 & \(19.534\) & \(0.046\) & \(0.890 \pm 0.002\) & \(0.579 \pm 0.001\) & \(130.3 \pm 0.1\) & E-M-A \\ 
SDSS\,J1537\(+\)0220 & \(19.682\) & \(0.111\) & \(0.386 \pm 0.001\) & \(0.694 \pm 0.001\) & \(\phantom{0}52.9 \pm 0.1\) & E-S-C \\ 
SDSS\,J1541\(+\)1812 & \(19.648\) & \(0.064\) & \(0.759 \pm 0.002\) & \(0.755 \pm 0.002\) & \(151.3 \pm 0.2\) & L-S-A \\ 
SDSS\,J1542\(+\)1629 & \(18.580\) & \(0.054\) & \(0.726 \pm 0.001\) & \(0.786 \pm 0.001\) & \(\phantom{0}89.5 \pm 0.1\) & E-S-A \\ 
SDSS\,J1545\(+\)2748 & \(18.594\) & \(0.050\) & \(2.589 \pm 0.005\) & \(0.661 \pm 0.001\) & \(105.2 \pm 0.1\) & E-S-A \\ 
SDSS\,J1601\(+\)2138 & \(19.712\) & \(0.127\) & \(0.436 \pm 0.001\) & \(0.960 \pm 0.001\) & \(110.5 \pm 1.0\) & E-S-A \\ 
SDSS\,J1611\(+\)1705 & \(19.611\) & \(0.085\) & \(0.998 \pm 0.002\) & \(0.926 \pm 0.002\) & \(\phantom{0}28.4 \pm 0.9\) & L-S-A \\ 
SDSS\,J1615\(+\)2056 & \(18.193\) & \(0.146\) & \(5.399 \pm 0.016\) & \(0.899 \pm 0.002\) & \(145.8 \pm 0.8\) & E-S-X \\ 
SDSS\,J1618\(+\)1930 & \(19.944\) & \(0.099\) & \(0.429 \pm 0.001\) & \(0.513 \pm 0.001\) & \(\phantom{00}9.2 \pm 0.1\) & L-S-X \\ 
SDSS\,J1627\(+\)1910 & \(19.477\) & \(0.089\) & \(0.360 \pm 0.000\) & \(0.661 \pm 0.001\) & \(101.6 \pm 0.1\) & E-S-X \\ 
SDSS\,J1631\(+\)1854 & \(17.859\) & \(0.079\) & \(1.434 \pm 0.001\) & \(0.929 \pm 0.001\) & \(166.7 \pm 0.3\) & E-S-A \\ 
SDSS\,J1637\(+\)1439 & \(19.147\) & \(0.093\) & \(1.037 \pm 0.002\) & \(0.674 \pm 0.001\) & \(122.9 \pm 0.1\) & L-S-A \\ 
SDSS\,J2122\(+\)0409 & \(19.501\) & \(0.147\) & \(0.903 \pm 0.002\) & \(0.819 \pm 0.002\) & \(136.6 \pm 0.3\) & E-S-A \\ 
SDSS\,J2125\(+\)0411 & \(18.271\) & \(0.150\) & \(0.901 \pm 0.001\) & \(0.689 \pm 0.001\) & \(111.0 \pm 0.1\) & E-S-A \\ 
SDSS\,J2303\(+\)0037 & \(18.721\) & \(0.081\) & \(1.349 \pm 0.002\) & \(0.760 \pm 0.001\) & \(\phantom{0}84.4 \pm 0.1\) & E-S-A  
 \enddata
    \tablecomments{Column 1 provides the SDSS System Name in terms of truncated J2000 R.A. and decl. in the
    format HHMM\(\pm\)DDMM.  Column 2 provides \textsl{HST}-ACS \(I_{814}\)-band apparent magnitudes based on flat-weighted de~Vaucouleurs models, and are quoted in the \(AB\) system~\protect\citep{Oke.1983ApJ...266..713O} without correction for Galactic extinction.  Column 3 provides \citet{Schlegel:1998ApJ...500..525S} Galactic extinction values that should be subtracted from the apparent magnitudes of Column 2 to give dust-corrected magnitudes.  Column 4 provides de~Vaucouleurs model intermediate axis effective radii. Column 5 provides axis ratios of minor to major axes, \(B/A\), for the de~Vaucouleurs models. Column 6 provides the major axis position angles for the de~Vaucouleurs models, measured east from north.  Column 7 provides our classification denoting (1) foreground-galaxy morphology, (2) foreground-galaxy multiplicity, and (3) status of system as a lens based on available data. Morphology is coded by ``E'' for early-type (elliptical and S0), ``L'' for late-type (Sa and later), and ``U'' for unclassified (galaxies that cannot be unambiguously classed as early- or late-type based on the \textsl{HST}-ACS \(I_{814}\)-band data).  Multiplicity is coded by ``S'' for single and ``M'' for multiple.  Lens grades, as justified in \tref{lensing:grade}, are coded by ``A'' for systems with clear and convincing evidence of multiple imaging, ``B'' for systems with strong evidence of multiple imaging but insufficient S/R for definite conclusion and/or modeling, ``C'' for systems with evidence of imaging such as single arcs or likely background features, but without detected counter-images, and ``X'' for all other systems (non-lenses).}
\end{deluxetable*}

To compute standardized model magnitudes, effective radii, projected axis ratios and position angles for the BELLS targets, we fit the images to two-dimensional ellipsoidal de~Vaucouleurs luminosity profiles. These fits are performed over a \(51\arcsec \times 51\arcsec\) square region centered on the target galaxies (approximately half the narrower dimension of the \textsl{HST} ACS-WFC CCD aperture in which the targets were roughly centered). The \(14\arcsec \times 14\arcsec\) masks created with our {\tt ACSPROC} software described in \iref{acsproc:mask} of \sref{acsproc} are applied in the central regions; stars and neighboring galaxies outside the manually masked area are masked with a single-step ``clipping'' of pixels that deviate by more than four standard deviations from the model. The fits are performed using the \texttt{MPFIT2DFUN} procedure in the {\tt IDL} programming language,  using a flat-weighted inverse variance, and include convolution with the appropriate rectified and stacked {\tt Tiny Tim} PSF~\protect\citep{Krist:1993ASPC...52..536K}.  The initial optimization is done by sampling the model at one point per data pixel, and repeated with a \(5 \times 5\) sub-sampling per pixel. 

The \(I_{814}\)-band apparent magnitudes are computed from the full (not truncated) analytic integral of the best-fit de~Vaucouleurs model, and provided in \tref{photmod}.  We also report the corrections for Galactic extinction from \citet{Schlegel:1998ApJ...500..525S}. Photometric stellar masses are computed from the extinction corrected \(I_{814}\)-band apparent magnitudes using the \citet{Maraston:2009MNRAS.394L.107M} LRG photometric stellar population model with a Salpeter initial mass function. We use the intermediate axis effective radius (the geometric mean of the major and minor axes).

\begin{deluxetable*}{lcp{14cm}}
    \tabletypesize{\scriptsize}
    \tablewidth{\hsize}
    \tablecaption{\label{table:lensing:grade}BELLS Confirmed Gravitational Lenses}
    \tablehead{
        \colhead{System Name} & 
        \colhead{Grade} & 
        \colhead{Justification} \\
        \colhead{\scriptsize (1)}&\colhead{\scriptsize (2)}&\colhead{\scriptsize (3)}
    }
    \tablecolumns{3}
    \startdata SDSS\,J0151\(+\)0049 & A & Simple single component source, lensed into a well modeled arc in the southest and a counter-image. \\ 
SDSS\,J0747\(+\)5055 & A & Image and counter-image are connected in partial Einstein ring, with an additional blob to the west. \\ 
SDSS\,J0747\(+\)4448 & A & Well modeled \(3+1\) quad with a faint but a definite counter-image, and a compact single source component. \\ 
SDSS\,J0801\(+\)4727 & A & Complex source structure with multiple arcs and counter-arcs. \\ 
SDSS\,J0830\(+\)5116 & A & Arc with a compact core to the northwest, and with a clear counter-image. \\ 
SDSS\,J0944\(-\)0147 & A & Well modeled arc and counter-arc, with a matching tail feature. \\ 
SDSS\,J1159\(-\)0007 & A & Well modeled arc and counter-image with detailed small scale features. \\ 
SDSS\,J1215\(+\)0047 & A & Well modeled double with detailed feature correspondence between the image and the counter-image. \\ 
SDSS\,J1221\(+\)3806 & A & Well modeled arc and counter-arc. The features are diffuse but significant. \\ 
SDSS\,J1234\(-\)0241 & A & Clear image and counter-image.  Possible foreground subtraction artifacts affect the counter-image in the north due to proximity to the lens center. Fixing the mass axis ratio to the light axis ratio increases model \(\theta_{E}\) by \(0\farcs03\). \\ 
SDSS\,J1318\(-\)0104 & A & Well modeled compact arc and counter-image. \\ 
SDSS\,J1337\(+\)3620 & A & Well modeled extended arc with a compact counter-image. \\ 
SDSS\,J1349\(+\)3612 & A & Well modeled arc and a compact counter-image.  Lens galaxy has a slightly disturbed morphology in its core. \\ 
SDSS\,J1352\(+\)3216 & A & Double image lens with detailed feature correspondence reproduced by the pixelized source plane model. \\ 
SDSS\,J1522\(+\)2910 & A & Double image lens with feature correspondence. \\ 
SDSS\,J1541\(+\)1812 & A & Well modeled compact double.  Diffuse component to the west reproduced by the pixelized model. \\ 
SDSS\,J1542\(+\)1629 & A & Well modeled arc and counter-arc. \\ 
SDSS\,J1545\(+\)2748 & A & Complex quad/double transition.  The detailed features are reproduced by the pixelized model. \\ 
SDSS\,J1601\(+\)2138 & A & Well modeled compact image and counter-image. \\ 
SDSS\,J1611\(+\)1705 & A & Double image lens with feature correspondence. \\ 
SDSS\,J1631\(+\)1854 & A & Spectacular high surface brightness Einstein ring with multiple source components. \\ 
SDSS\,J1637\(+\)1439 & A & Well modeled arc and counter-arc. \\ 
SDSS\,J2122\(+\)0409 & A & Well modeled arc and counter-arc with individual knots that are reproduced by the pixelized source model. \\ 
SDSS\,J2125\(+\)0411 & A & Extended arc and counter-arc with detailed substructure reproduced by the pixelized source model. \\ 
SDSS\,J2303\(+\)0037 & A & Clear \(3+1\) quad morphology.  There is a galaxy group $10\arcsec$ to the east that may contribute shear at the lens position and account for the model short comings.  
\\ \tableline \\ SDSS\,J0915\(-\)0055 & B & Plausibly modeled image and counter-image, including substructure reproduced by the pixelized source model.  Classified Grade-B because of the low surface brightness and the significant bright star light contamination in the south. \\ 
SDSS\,J1452\(+\)3323 & B & Very faint possible Einstein-ring features, but the surface brightness and S/R are too low for modeling. \\ 
SDSS\,J1503\(+\)3225 & B & Very plausible image-counter-image pair to the northwest and southeast, with detailed feature correspondence, but the group environment of the lens complicates modeling and interpretation. \\ 
\\ \tableline \\ SDSS\,J0821\(+\)3733 & C & Likely background features seen in the residual image, but no counter-images. \\ 
SDSS\,J0840\(+\)5051 & C & Likely background features seen in the residual image, but no counter-images. \\ 
SDSS\,J0841\(+\)5017 & C & Extended source to south inconsistent with the compact source to the north as image-counter-image pair. \\ 
SDSS\,J0941\(-\)0104 & C & Clear distorted arc, but no detected counter-image. \\ 
SDSS\,J1016\(-\)0208 & C & Tangentially elongated image to the northeast with no detected counter-image. \\ 
SDSS\,J1039\(-\)0014 & C & Disky late-type structure with lensed arc to the east, but no apparent counter-image. \\ 
SDSS\,J1344\(+\)3258 & C & Clear evidence of background-galaxy to the west, but no evidence of a counter-image. \\ 
SDSS\,J1537\(+\)0220 & C & Pronounced asymmetry in the lens galaxy brightness profile; plausible background-galaxy arc to the southwest, but no evidence of a counter-image. \\ 
 \enddata
    \tablecomments{Column 1 provides the SDSS System Name in terms of truncated J2000 R.A. and decl. in the
    format HHMM\(\pm\)DDMM.  Column 2 provides our lens-grade, coded by ``A'' for the 25 systems with clear and convincing evidence of multiple imaging, ``B'' for the 3 systems with strong evidence of multiple imaging but insufficient S/R for definite conclusion and/or modeling, ``C'' for the 8 systems with evidence of imaging such as single arcs or likely background features, but without detected counter-images.  Column 3 provides our lens-grade justification. SIE model parameters for the grade-A lenses are listed in \tref{lensing:parameters}. The remainder of the observed sample, listed in \tref{appendix:gradex}, receives a grade of ``X'' (non-lenses).}
\end{deluxetable*}

\fref{population.magnitude} shows that the distribution of grade-A lenses in \textsl{HST} \(I_{814}\)-band magnitudes is similar to that of the SLACS sample, lying at higher redshifts, but within the same stellar mass range of \(10^{11}M_{\sun} \lesssim M_\star \lesssim 10^{12}M_{\sun}\).  The distribution in the ratio \(\theta_E/R_{\rm eff}\) of the Einstein radius to the lens effective radius as a function of the lens redshift is shown in \fref{population.size}.  The BELLS sample extends the SLACS sample to a wider range of \(\theta_{E}/R_{\rm eff}\) while still providing substantial overlap with the range of ratios seen in the SLACS sample.

Each of the \textsl{HST}-ACS \(I_{814}\)-band images was visually inspected to classify all targets for multiplicity and morphology.  Systems with two or more foreground-galaxies of comparable luminosity are classified as ``multiple'', while systems with only a single dominant foreground-galaxy are classified as ``single''.  Morphological classification is limited to the categories of ``early-type'' (elliptical and S0), ``late-type'' (Sa and later spirals), and ``unclassified'' (generally ambiguous between S0 and Sa).  The classification of observed candidates into lenses and non-lenses is made by visual examination of the direct and B-spline model-subtracted residuals, based on the appearance of arcs, rings, and multiple images centered on the position of the foreground-galaxy.  Our classification is provided in the \tref{photmod}, with our justifications for BELLS confirmed lenses provided in \tref{lensing:grade}.

\section{Strong Lensing Analysis} \label{section:lensing} \figuretitle{SIE Lens Models with S\'ersic Source Models as compared to Pixelized Source Models}
\begin{figure*}[t]
    \plotone{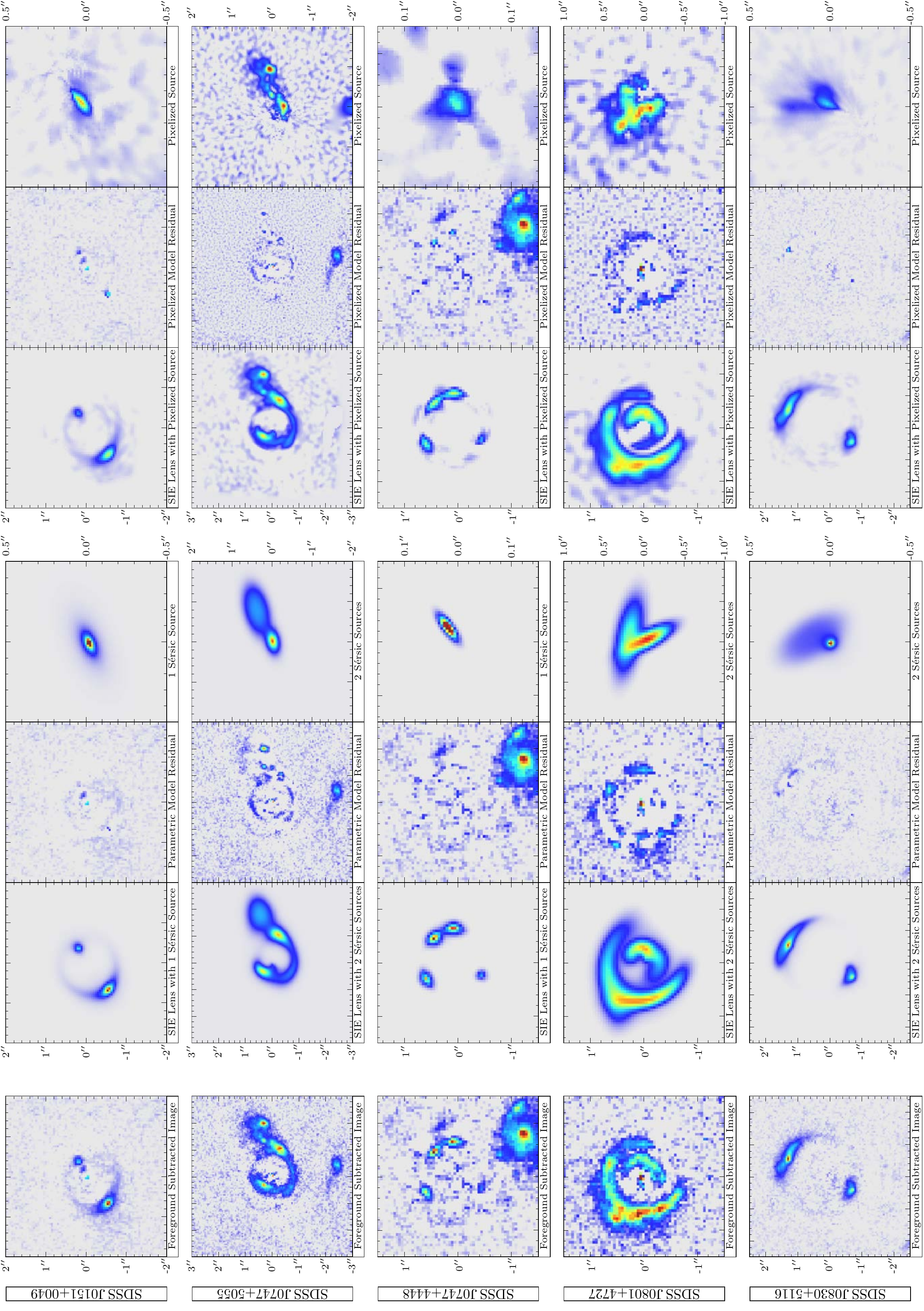} 
    \caption{\label{figure:lensing:models}\thefiguretitle. SIE Lens Models are shown for each of the 25 confirmed-to-date Grade-A lenses discovered under \textsl{HST} Cycle 18 Program ID GO-12209.  For each system, the left panel shows the lensed features which remain after foreground-galaxy subtraction by the B-spline method described in \sref{acsproc}, repeated from the rightmost panel of \fref{acsproc}. Best-fit SIE Models with S\'ersic sources are shown in the second panel from the left, including the residuals of the data minus the parametric model in the third panel, with the de-lensed S\'ersic source model shown in the fourth panel, at a scale optimized for the source plane.  Best-fit SIE Model parameters are provided in \tref{lensing:parameters}. SIE Models with pixelized sources are shown in the fifth panel from the left, including the residuals of the data minus the pizelized model in the sixth panel, with the de-lensed pixelized source model shown in the rightmost panel, at the same scale as used for the parametric source model.  Comments justifying our lens grade are provided in \tref{lensing:grade}. The color scale is provided in \fref{acsproc}. This figure is \emph{continued}.}
\end{figure*} \addtocounter{figure}{-1}
\begin{figure*}[p] \plotone{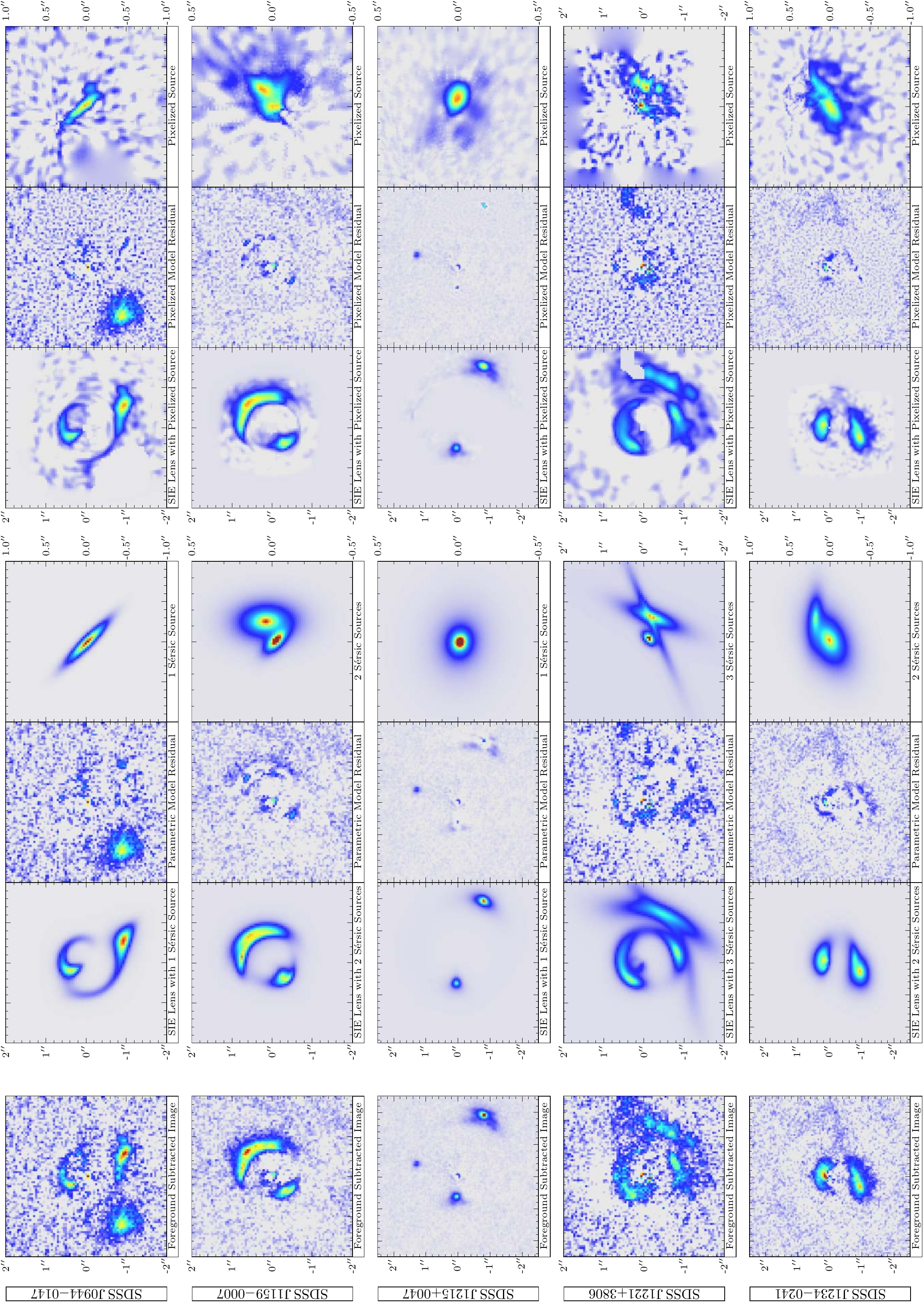} \caption{\emph{Continued}. \thefiguretitle.}\end{figure*} \addtocounter{figure}{-1}
\begin{figure*}[p] \plotone{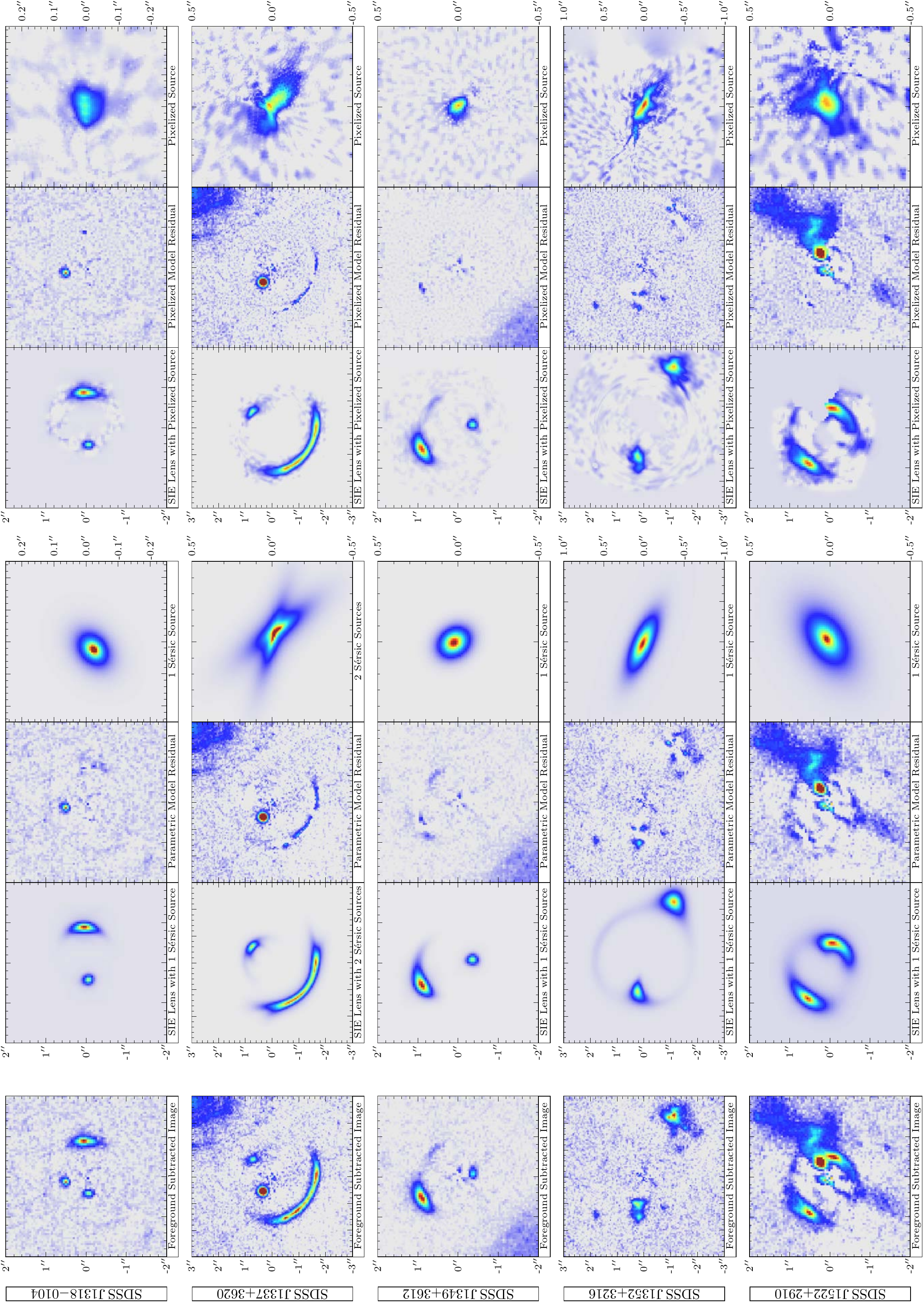} \caption{\emph{Continued}. \thefiguretitle.}\end{figure*} \addtocounter{figure}{-1}
\begin{figure*}[p] \plotone{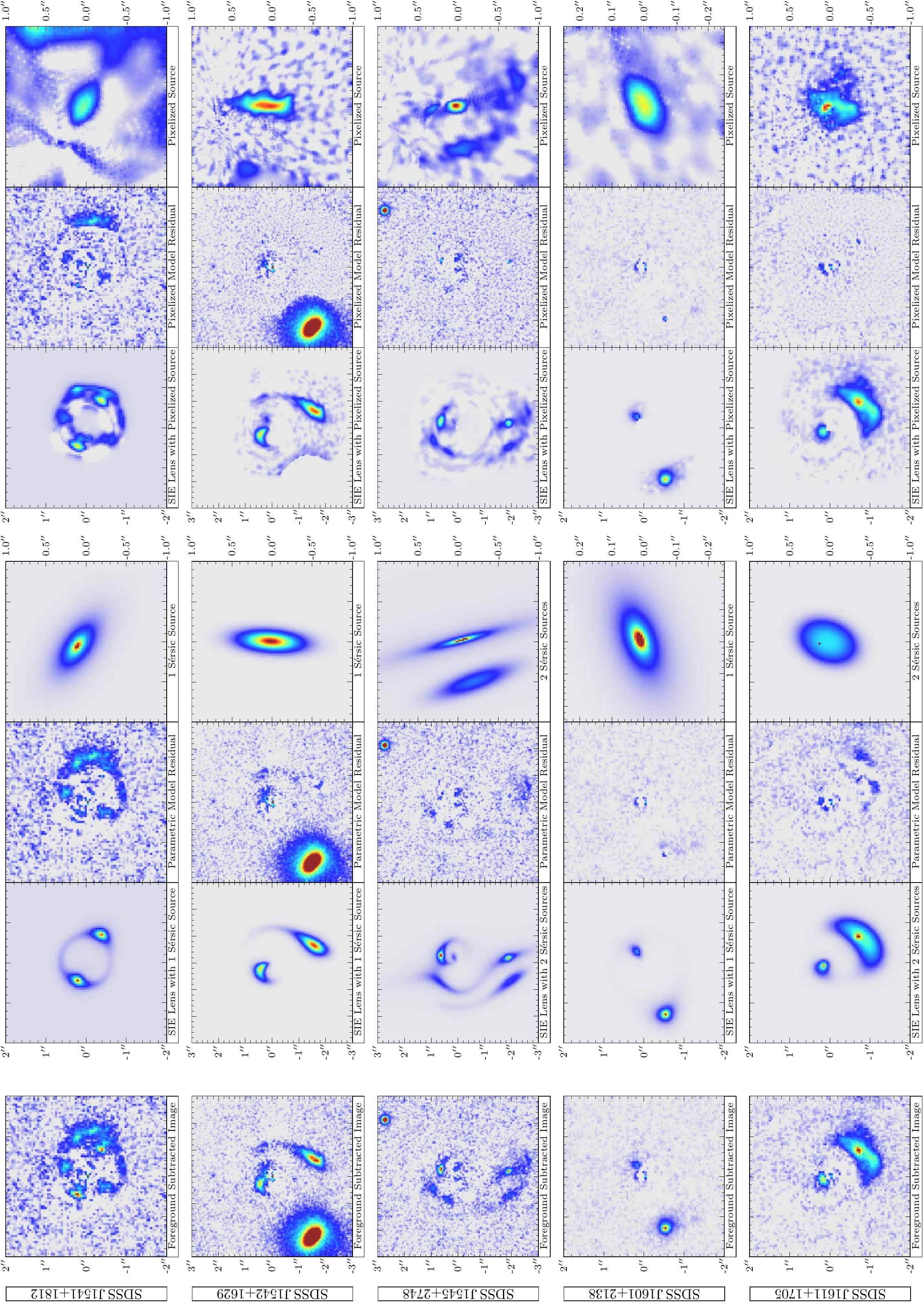} \caption{\emph{Continued}. \thefiguretitle.}\end{figure*} \addtocounter{figure}{-1}
\begin{figure*}[p] \plotone{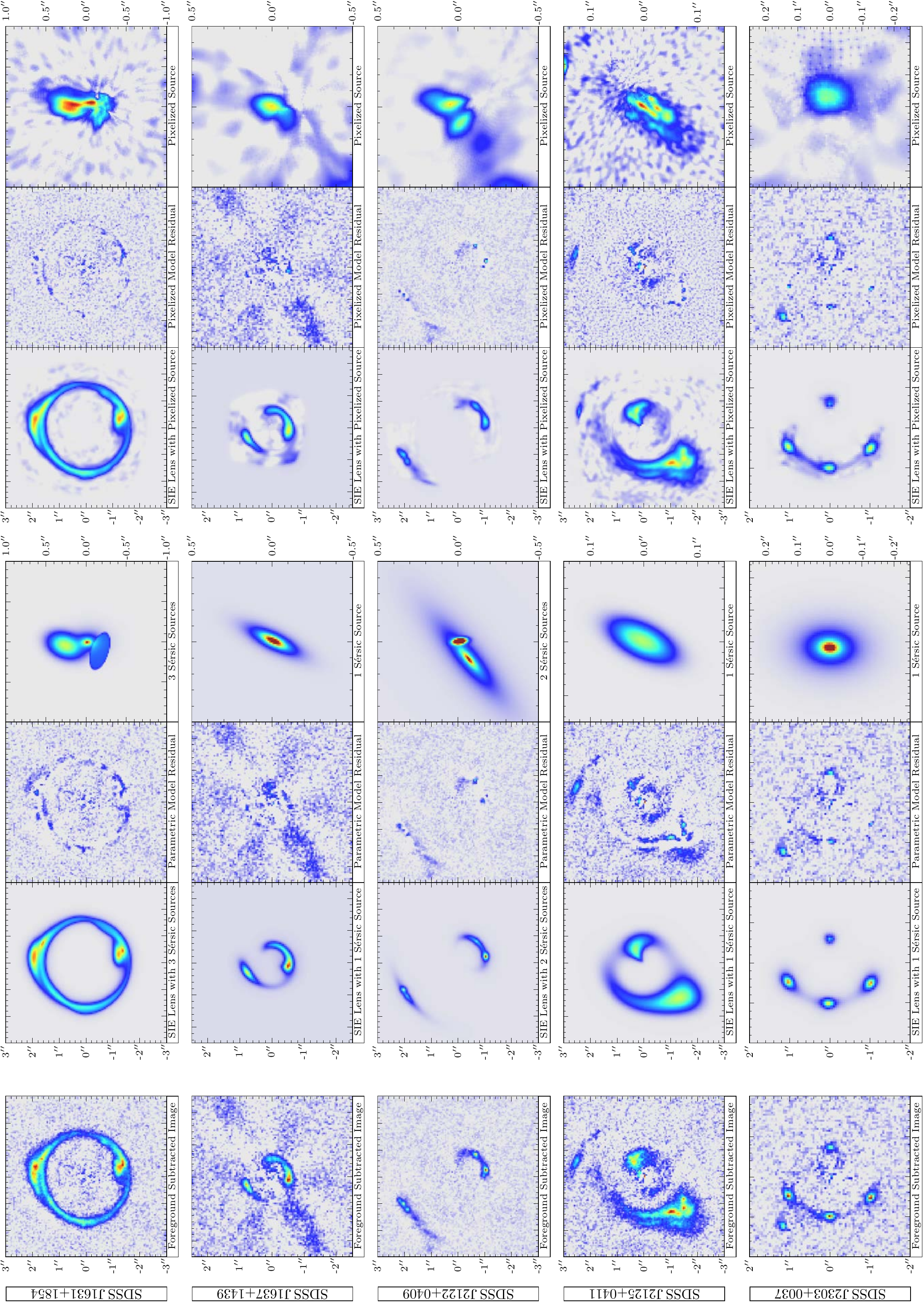} \caption{\emph{Continued}. \thefiguretitle.}\end{figure*}

\fref{acsproc} shows the foreground-subtracted images of each BELLS target that completed the {\tt ACSPROC} image reduction process described in \sref{acsproc}.  These are the lensed features which remain after the foreground-galaxy has been subtracted.  This image is passed forward to the lens modeling stage as the zeroth extension of a {\tt BIZZLE} {\tt FITS} file.  In addition, the inverse variance of the foreground-subtracted image, the lensed features mask, the mask for extraneous unlensed features, and the \textsl{HST} ACS-WFC PSF are included in successive {\tt FITS} extensions.  The best-fit B-spline model parameters of the foreground-galaxy are included in the {\tt FITS} header so that the center of the lens model may be constrained by the luminous distribution, and so that the ellipticity of the visible galaxy may be used to initialize the lens model.  These BELLS {\tt BIZZLE} files are input into our {\tt ACSLENS} software, implemented in the {\tt Python} programming language, which accommodates the SIE lens model~\protect\citep{Kassiola.1993ApJ...417..450K,Kormann.1994AA...284..285K,Keeton.1998ApJ...495..157K}, and both parametric or pixelized source models.  

\begin{deluxetable}{cccrcrr}
    \tabletypesize{\scriptsize}
    \tablewidth{\hsize}
    \tablecaption{\label{table:lensing:parameters}BELLS Grade-A Strong Lens SIE Model Parameters}
    \tablehead{
        \colhead{System Name} & 
        \colhead{\(\theta_E\) (\(\arcsec\))} & 
        \colhead{\(q_{\mathrm{SIE}}\)} & 
        \colhead{\(P.A.\) (\(\degr\))} & 
        \colhead{\(N_S\)} &
        \colhead{\(m_{814}\)} &
        \colhead{\(\mu\)} \\
        \colhead{\scriptsize (1)}&\colhead{\scriptsize (2)}&\colhead{\scriptsize (3)}&\colhead{\scriptsize (4)}&\colhead{\scriptsize (5)}&\colhead{\scriptsize (6)}&\colhead{\scriptsize (7)}
    }
    \tablecolumns{7}
    \startdata SDSS\,J0151\(+\)0049 & \(0.676\) & \(0.752\) & \(111.0\) & \(1\) & \(22.51\)  & \(8.71\) \\ 
SDSS\,J0747\(+\)5055 & \(0.754\) & \(0.641\) & \(4.9\) & \(2\) & \(21.46\)  & \(2.95\) \\ 
SDSS\,J0747\(+\)4448 & \(0.610\) & \(0.723\) & \(147.1\) & \(1\) & \(23.77\)  & \(39.72\) \\ 
SDSS\,J0801\(+\)4727 & \(0.491\) & \(0.891\) & \(41.1\) & \(2\) & \(22.07\)  & \(3.82\) \\ 
SDSS\,J0830\(+\)5116 & \(1.142\) & \(0.887\) & \(137.0\) & \(2\) & \(21.89\)  & \(7.43\) \\ 
SDSS\,J0944\(-\)0147 & \(0.725\) & \(0.922\) & \(108.4\) & \(1\) & \(23.29\)  & \(5.09\) \\ 
SDSS\,J1159\(-\)0007 & \(0.683\) & \(0.815\) & \(125.3\) & \(2\) & \(21.92\)  & \(9.46\) \\ 
SDSS\,J1215\(+\)0047 & \(1.368\) & \(0.742\) & \(123.2\) & \(1\) & \(21.25\)  & \(3.69\) \\ 
SDSS\,J1221\(+\)3806 & \(0.699\) & \(0.745\) & \(91.4\) & \(3\) & \(22.50\)  & \(3.43\) \\ 
SDSS\,J1234\(-\)0241 & \(0.533\) & \(0.279\) & \(77.7\) & \(2\) & \(21.92\)  & \(2.56\) \\ 
SDSS\,J1318\(-\)0104 & \(0.679\) & \(0.836\) & \(105.2\) & \(1\) & \(22.87\)  & \(6.13\) \\ 
SDSS\,J1337\(+\)3620 & \(1.386\) & \(0.682\) & \(137.4\) & \(2\) & \(21.71\)  & \(11.55\) \\ 
SDSS\,J1349\(+\)3612 & \(0.750\) & \(0.711\) & \(177.5\) & \(1\) & \(22.06\)  & \(4.79\) \\ 
SDSS\,J1352\(+\)3216 & \(1.823\) & \(0.859\) & \(115.2\) & \(1\) & \(21.97\)  & \(5.65\) \\ 
SDSS\,J1522\(+\)2910 & \(0.736\) & \(0.741\) & \(143.7\) & \(1\) & \(22.24\)  & \(5.55\) \\ 
SDSS\,J1541\(+\)1812 & \(0.640\) & \(0.927\) & \(142.4\) & \(1\) & \(23.80\)  & \(9.76\) \\ 
SDSS\,J1542\(+\)1629 & \(1.042\) & \(0.806\) & \(105.6\) & \(1\) & \(22.32\)  & \(3.28\) \\ 
SDSS\,J1545\(+\)2748 & \(1.210\) & \(0.418\) & \(99.0\) & \(2\) & \(22.58\)  & \(4.35\) \\ 
SDSS\,J1601\(+\)2138 & \(0.858\) & \(0.906\) & \(164.6\) & \(1\) & \(22.84\)  & \(3.50\) \\ 
SDSS\,J1611\(+\)1705 & \(0.580\) & \(0.740\) & \(4.3\) & \(2\) & \(22.13\)  & \(2.27\) \\ 
SDSS\,J1631\(+\)1854 & \(1.634\) & \(0.878\) & \(35.3\) & \(3\) & \(20.41\)  & \(19.15\) \\ 
SDSS\,J1637\(+\)1439 & \(0.650\) & \(0.750\) & \(122.5\) & \(1\) & \(22.99\)  & \(10.65\) \\ 
SDSS\,J2122\(+\)0409 & \(1.580\) & \(0.634\) & \(119.7\) & \(2\) & \(22.31\)  & \(5.73\) \\ 
SDSS\,J2125\(+\)0411 & \(1.204\) & \(0.821\) & \(83.0\) & \(1\) & \(21.11\)  & \(4.32\) \\ 
SDSS\,J2303\(+\)0037 & \(1.016\) & \(0.391\) & \(88.8\) & \(1\) & \(23.10\)  & \(8.06\)  
 \enddata
    \tablecomments{The method used to compute the best-fit SIE model parameters is described in \sref{lensing}. Column 1 provides the SDSS System Name in terms of truncated J2000 R.A. and decl. in the
    format HHMM\(\pm\)DDMM.  Column 2 provides the best-fit SIE model Einstein radius.  Column 3 provides the best-fit minor to major  axis ratio for the SIE models. Column 4 provides the major axis position angle for the SIE models, measured east from north.  Column 5 provides the number of S\'ersic models added to the source plane to provide the best-fit.  Column 6 provides the \(I_{814}\)-band magnitude of the lensed source.  Column 7 provides the magnification of the lensed source.}
\end{deluxetable}

 The SIE model is a two-dimensional potential of similar concentric and aligned elliptical isodensity contours, with axis ratio \(q_{\mathrm{SIE}}\). The projected surface density of the SIE falls off as \(1/R\), and the model provides flat galaxy rotation curves. The model is parameterized by its angular Einstein radius, \(\theta_{E}\), which is related to the physical mass model through
 \begin{equation}\label{equation.einsteinradius}
 \theta_{E} = 4 \pi {{\sigma_{\mathrm{SIE}}^2} \over {c^2}} {D_{LS} \over D_S},
 \end{equation}
 where \(\sigma_{\mathrm{SIE}}\) is a velocity dispersion parameter and \(D_{LS}\) and \(D_S\) are cosmological angular diameter distances from lens and observer to source, respectively. We adopt the intermediate-axis normalization of \citet{Kormann.1994AA...284..285K}, whereby the mass within a given isodensity contour remains constant at fixed \(\theta_{E}\) for changing \(q_{\mathrm{SIE}}\).

We attempt to model all candidate lenses with an SIE lens-mass model and a parametric source-plane surface-brightness model, where the source may be composed of either single or multiple Gaussian, \citet{deVaucouleurs.1948AnAp...11..247D} or \citet{Sersic.1968adga.book.....S} distributions, as necessary to obtain a good (\(\chi^2\)) fit.  The model lensed image is generated by ray-tracing through the analytic SIE mass model, and then convolved with the \textsl{HST} ACS-WFC PSF\@.  After choosing reasonable initial parameter values, the SIE and source-plane model parameters are optimized nonlinearly with the \texttt{leastsq} Levenberg--Marquardt implementation in the \texttt{scipy.optimize} package.  In the SIE mass-model optimization, the mass centroid is fixed to the centroid of the luminous distribution as determined from the B-spline lens-galaxy photometric analysis, while the SIE strength, axis ratio, and major-axis position angle are all allowed to vary.  The final outcome is a set of lens-model and source-component parameters, along with a model for the lensed images.

Our best-fit parametric source model is then complemented with a pixelized source-plane model, which we compute through a matrix inversion including linear regularization, without re-optimizing the lens model~\protect\citep{Warren.2003ApJ...590..673W,Treu.2004ApJ...611..739T,Wayth.2006MNRAS.372.1187W,Koopmans:2006ApJ...649..599K}.

If necessary to obtain a better correspondence between the parametric and pixelized model images, the apparent multiplicity and spatial orientation of the pixelized source distribution are used to place additional S\'ersic sources, and then the parametric model is re-optimized together with the SIE mass-model parameters.  Several iterations are pursued until the parametric and pixelized models are in reasonable agreement.
Although we do not use the pixelized models in the lens model optimization, their ability to capture more
detail in the lensed galaxies serves as an important confirmation of the lensing hypothesis in most cases.

Our best-fit SIE lens models with S\'ersic source models, along with our pixelized source models, are presented in \fref{lensing:models}, and the best-fit model parameters are listed in \tref{lensing:parameters}.  Our justifications for the BELLS confirmed lenses grading are provided in \tref{lensing:grade}.

\section{Summary and Conclusions} \label{section:conclusions} We present 44 of the 45 targets observed-to-date in the BOSS Emission-Line Lens Survey (BELLS) from the \textsl{HST} ACS-WFC Cycle 18 Program ID GO-12209 imaging program.  The BELLS catalog presently includes 25 grade-A strong galaxy--galaxy lenses complete with lens and source redshifts, \(I_{814}\)-band foreground-galaxy photometry, and SIE models.  This represents a significant expansion of the SLACS survey, extending the lens redshifts to  \(z_{\mathrm{L}} \simeq 0.66\), and with source redshifts extended to \(z_{\mathrm{S}}\simeq 1.52\).  The combined SLACS and BELLS catalog provides a homogeneous sample of strong lens galaxies at significant cosmological look-back times.  Future papers in this series will apply the combined SLACS and BELLS samples to the analysis of the structure and evolution of massive elliptical galaxies.

We have described the use of our {\tt ACSPROC} \textsl{HST} ACS-WFC image processing software to make luminosity measurements and to produce our {\tt BIZZLE} data product.  We then analyze the {\tt BIZZLE} data using our {\tt ACSLENS} lens-modeling software to make parameterized and pixelized lens models. Our analysis demonstrates that simple SIE lens models, combined with multiple S\'{e}rsic ellipsoid models of the lensed background-galaxies, can reproduce the lensed image configurations in great detail, and our pixelized source models reproduce more detailed features in the lensed images and add finer structure to the de-lensed source model.

The high success rate of our \textsl{HST} imaging program demonstrates that the spectroscopic strong-lens selection methodology can be successfully applied to the BOSS spectroscopic database even though BOSS is targeting a more distant --- and hence fainter --- galaxy population than SDSS-I\@.  The probability of finding a galaxy--galaxy lens using the spectroscopic discovery method was considered in \citet{Dobler:2008ApJ...685...57D} for the SLACS survey, including selection effects such as the finite size of the spectroscopic fiber (which selects against large separation lenses), and the effectiveness of spectral noise modeling (which selects against sources that have redshifted emission-lines coincident with strong emission-lines in the sky). The probability that a galaxy--galaxy lens would be spectroscopically discovered from either the SDSS-I or SDSS-III spectra, and lead to a confirmed lens, is under consideration in \citet{Arneson:2011} using a Monte Carlo simulation to measure the effects of variation in the lens mass, ellipticity and morphology, source extent, and the logarithm slope of the lens luminosity profile.

The strong lensing measurements presented in this work afford a unique opportunity to test the results of theoretical models and numerical simulations of galaxy formation, merging, and evolution. This is due to the fact that strong lensing measures the projected mass inside the aperture defined by the Einstein radius, in a nearly model-independent sense, and without need for modeling of stellar populations and luminosity evolution.

In comparison with the SDSS-I spectra of the SLACS lens sample, the BOSS spectra for the BELLS sample have significantly lower signal to noise.  Hence, the resulting velocity dispersion measurements for the BELLS sample are correspondingly noisier and more care must be taken in interpreting these measurements in the context of dynamical galaxy modeling.  Work in preparation will present a joint analysis of all BELLS lenses that marginalizes over the full velocity dispersion likelihood function for individual spectra \citep[see][]{Shu.2011arXiv1109.6678S}.  Deeper spectroscopy with large ground-based telescopes can provide more precise stellar velocity dispersion measurements.

We note that the BOSS spectroscopic database as of summer 2011, which will constitute the ninth SDSS public Data Release (DR9), contains 819 unique BOSS plates
with \pagebreak an average of approximately 600 successfully redshifted galaxies per plate.  (The increase in the number of spectroscopically confirmed galaxies per plate was produced by a combination of improvements in targeting, hardware, operations, and analysis software during the BOSS commissioning phase and early operations.) Thus, DR9 is at least a fourfold increase over the sample we used here.  The final BOSS sample, to be completed in mid-2014 and released at the end of that year, will be an order of magnitude increase over the sample used here.  In addition, ongoing BOSS spectroscopic pipeline development (including implementation of the two dimensional PSF-based extraction algorithm of \citealt{Bolton.2010PASP..122..248B}) continues to increase the quality of the reduced spectra and redshift/classification analysis for BOSS significantly.  Hence, there will be at least several hundred strong lenses to be found in BOSS spectroscopy at the end of the survey, offering the possibility of measuring the cosmic evolution of the relationship between mass-density structure and stellar mass, luminosity, and rest-frame color, in great quantitative detail.
\acknowledgements{
Support for program 12209 was provided by NASA through a grant from the Space Telescope Science Institute, which is operated by the Association of Universities for Research in Astronomy, Inc., under NASA contract NAS 5-26555. This work has made extensive use of the Baryon  Oscillation Spectroscopic Survey, part of the Sloan Digital Sky Survey III (SDSS-III). Funding for SDSS-III has been provided by the Alfred P. Sloan Foundation, the Participating Institutions, the National Science Foundation, and the U.S. Department of Energy Office of Science. The SDSS-III Web site is \url{http://www.sdss3.org/}. SDSS-III is managed by the Astrophysical Research Consortium for the Participating Institutions of the SDSS-III Collaboration including the University of Arizona, the Brazilian Participation Group, Brookhaven National Laboratory, University of Cambridge, University of Florida, the French Participation Group, the German Participation Group, the Instituto de Astrofisica de Canarias, the Michigan State/Notre Dame/JINA Participation Group, Johns Hopkins University, Lawrence Berkeley National Laboratory, Max Planck Institute for Astrophysics, New Mexico State University, New York University, Ohio State University, Pennsylvania State University, University of Portsmouth, Princeton University, the Spanish Participation Group, University of Tokyo, University of Utah, Vanderbilt University, University of Virginia, University of Washington, and Yale University.  J.R.B. and A.S.B. acknowledge the hospitality of the Max-Planck-Institut f\"{u}r Astronomie, where a portion of this work was completed.  C.S.K. is supported by NSF grant AST-1004756.
\vspace*{0.0mm}
}
\bibliographystyle{hapj}
\bibliography{paper}
\setcounter{figure}{0}
\setcounter{table}{0}
\renewcommand{\thefigure}{A\arabic{figure}}
\renewcommand{\thetable}{A\arabic{table}}
\onecolumngrid
\figuretitle{BELLS Grade-X Non-Lens Galaxies}
\begin{figure*}[t]
\begin{center}
    \vspace*{0.125cm}
    \noindent\mbox{}\hfill{\apjsecfont APPENDIX}\hfill\mbox{}\par
    \vspace*{0.5cm}
    \includegraphics[width=0.8\textwidth]{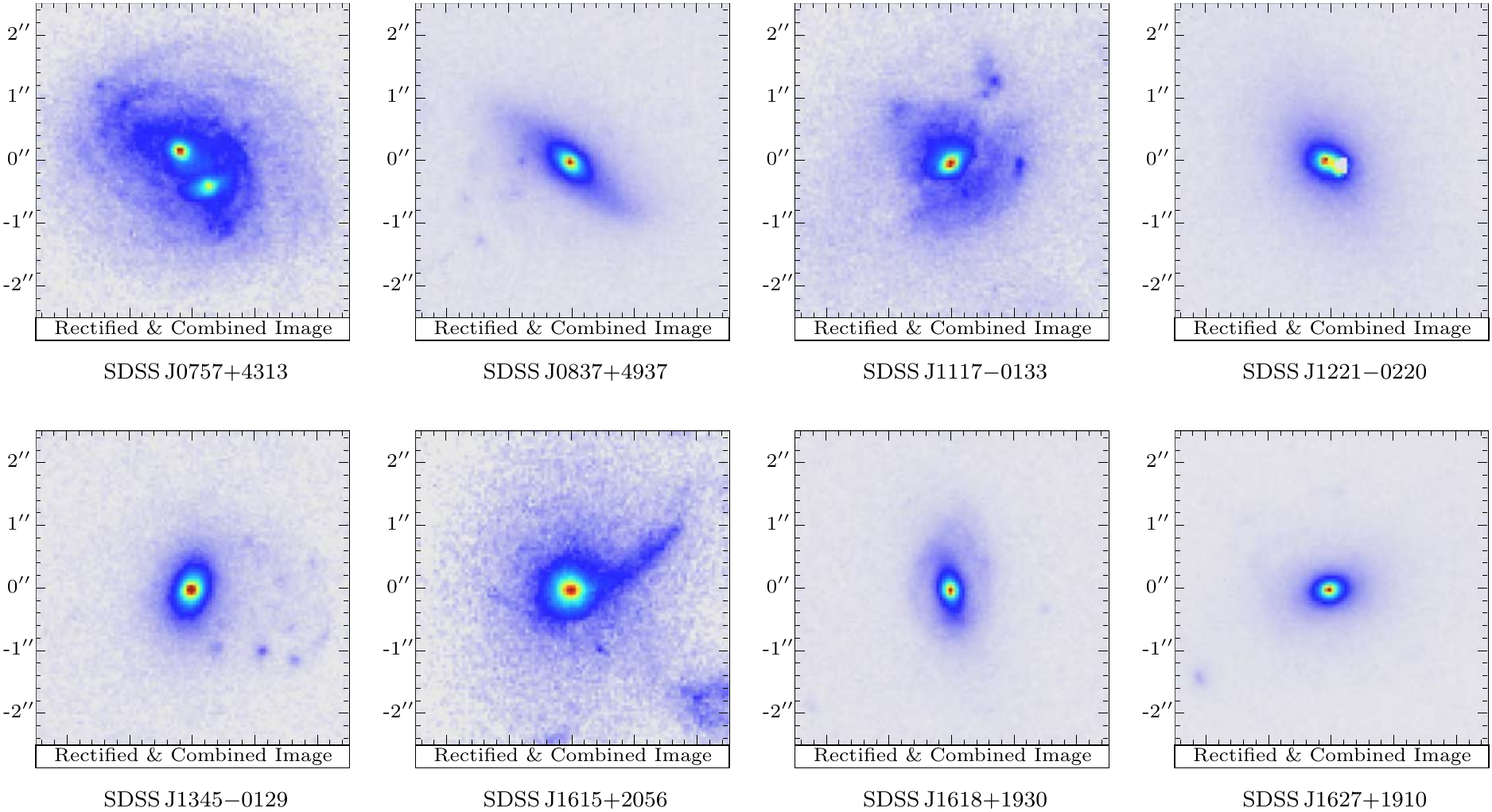} 
    \caption{\label{figure:appendix:GradeX}\thefiguretitle. The 8 non-lenses, discovered to-date under \textsl{HST} Cycle 18 Program ID GO-12209, \tref{systems} presents their properties as derived from the BOSS data, and \tref{photmod} presents those derived from the \textsl{HST} data. For each system, the panel shows the \textsl{HST} ACS-WFC F814W rectified and combined images with north up and east to the left.  Comments justifying the lens-grade of ``X'' are provided in \tref{appendix:gradex}. The color scale is provided in \fref{acsproc}.}
\end{center}
\end{figure*}
\begin{deluxetable*}{lp{15cm}}
    \tabletypesize{\scriptsize}
    \tablewidth{\hsize}
    \tablecaption{\label{table:appendix:gradex}BELLS Grade-X Non-Lens Galaxies}
    \tablehead{
        \colhead{System Name} & 
        \colhead{Justification} \\
        \colhead{\scriptsize (1)}&\colhead{\scriptsize (2)}
    }
    \tablecolumns{2}
    \startdata SDSS\,J0757\(+\)4313 & Two components with spiral structure; any lensing features are difficult to disentangle. \\ 
SDSS\,J0837\(+\)4937 & Late-type with faint candidate features, but with no evidence of multiple imaging. \\ 
SDSS\,J1117\(-\)0133 & Pronounced asymmetry in the foreground-galaxy plus possible spiral and dust features. Possible lensed background sources cannot be unambiguously modeled given single-band imaging. \\ 
SDSS\,J1221\(-\)0220 & Double nucleus in the foreground-galaxy, with no evidence of background-galaxy features. \\ 
SDSS\,J1345\(-\)0129 & Faint possible background-galaxy features to the west, but without plausible evidence of strong lensing. \\ 
SDSS\,J1615\(+\)2056 & Possible candidate-arc extending to the northwest and to the east, but the system is difficult to interpret as a lens. \\ 
SDSS\,J1618\(+\)1930 & No obvious background features, but spiral structure and/or dust in the foreground-galaxy complicates interpretation. \\ 
SDSS\,J1627\(+\)1910 & Possible faint background-galaxy features to the northeast, but without evidence of strong lensing. \\ 
 \enddata
    \tablecomments{Column 1 provides the SDSS System Name in terms of truncated J2000 R.A. and decl. in the
    format HHMM\(\pm\)DDMM.  Column 2 provides our justification for the lens-grade of ``X'' (non-lenses).}
\end{deluxetable*}
\end{document}